%% file: main.tex
\PassOptionsToPackage{dvipsnames}{xcolor}
\documentclass[11pt]{article}

\usepackage[preprint]{acl}

\usepackage{times}
\usepackage{latexsym}
\usepackage[T1]{fontenc}
\usepackage[utf8]{inputenc}
\usepackage{microtype}
\usepackage{inconsolata}

\usepackage{amsmath}
\usepackage{amsfonts}
\usepackage{algorithm}
\usepackage{algorithmic}
\usepackage{booktabs}
\usepackage{enumitem}
\usepackage{graphicx}
\usepackage{multirow}
\usepackage{pifont}
\usepackage{siunitx}
\usepackage{tcolorbox}

\tcbuselibrary{skins,breakable}

\newcommand{\cmark}{\ding{51}}
\newcommand{\best}[1]{\textbf{#1}}
\newcommand{\second}[1]{\underline{#1}}
\newcommand{\tok}[1]{\num{#1}}

\title{RePair: Turning Retrieval Failures into Counterfactual Hard Pairs}

\author{%
  \bfseries
  Siyi Liu\textsuperscript{1,}\footnotemark[2]\quad
  Xiaorong Zhu\textsuperscript{1,}\footnotemark[2]\quad
  Enjun Du\textsuperscript{1,2,3}\quad
  Xinyu Zuo\textsuperscript{2}\quad
  Lisheng Duan\textsuperscript{2} \\
  \bfseries
  Haijin Liang\textsuperscript{2}\quad
  Jin Ma\textsuperscript{2}\quad
  Junfu Pu\textsuperscript{4}\quad
  Yongqi Zhang\textsuperscript{1,}\footnotemark[1] \\[0.5ex]
  \normalfont\normalsize
  \textsuperscript{1}The Hong Kong University of Science and Technology (Guangzhou) \\
  \normalfont\normalsize
  \textsuperscript{2}Tencent Yuanbao\qquad
  \textsuperscript{3}The University of Hong Kong\qquad
  \textsuperscript{4}ARC Lab, Tencent \\
  \normalfont\normalsize
  \texttt{ssui.liu1022@gmail.com}\qquad
  \texttt{xrzhulalala@163.com} \\
  \normalfont\normalsize
  \texttt{yongqizhang@hkust-gz.edu.cn}
}

\begin{document}
\maketitle
\begingroup
  \renewcommand{\thefootnote}{\fnsymbol{footnote}}
  \footnotetext[2]{Equal contribution.}
  \footnotetext[1]{Corresponding author.}
\endgroup

\begin{abstract}
\input{sections/abstract}
\end{abstract}

\input{sections/introduction}
\input{sections/related-work}
\input{sections/method}
\input{sections/experiments}
\input{sections/conclusion}

\section*{Limitations}
\input{sections/limitations}

\section*{Ethics Statement}
\input{sections/ethics-statement}

\section*{Acknowledgments}
\input{sections/acknowledgments}

\bibliography{references}

\section*{Code Availability}
Our code is available at: \url{https://github.com/ssui-liu/RePair}.

\appendix
\input{sections/appendix}

\end{document}

%% file: sections/abstract.tex
Vision-language retrieval with CLIP-style dual encoders achieves strong cross-modal performance, yet practical accuracy often hinges on localized semantic distinctions where top-ranked near misses differ from the true match by a single critical detail. Hard-sample mining can select confusable candidates but cannot construct corrected counterparts; synthetic augmentation can generate novel samples but, without conditioning on actual model failures, targets irrelevant dimensions of hardness. We observe that a top-ranked false positive is a counterfactual scaffold---sharing most of the query's semantics while differing in a localized failure-causing residual. Minimally correcting this residual yields a hard positive of the ground truth in the same modality; the corrected and unedited versions form a hard negative pair that straddles the decision boundary, producing complementary pull--push supervision. We introduce RePair, guided by three principles---Validity, Minimality, and Locality---which mines false positives bidirectionally, applies LLM-guided counterfactual editing, and trains with a local hard-pair contrastive objective. On Flickr30K and COCO30K, RePair outperforms controlled augmentation baselines with only 107K synthetic samples---26\%--75\% fewer than comparable methods---confirming failure-conditioned repair is more data-efficient than error-agnostic augmentation.

%% file: sections/introduction.tex
\section{Introduction}
\label{sec:introduction}

\begin{figure}[t]
\centering
\includegraphics[width=\columnwidth]{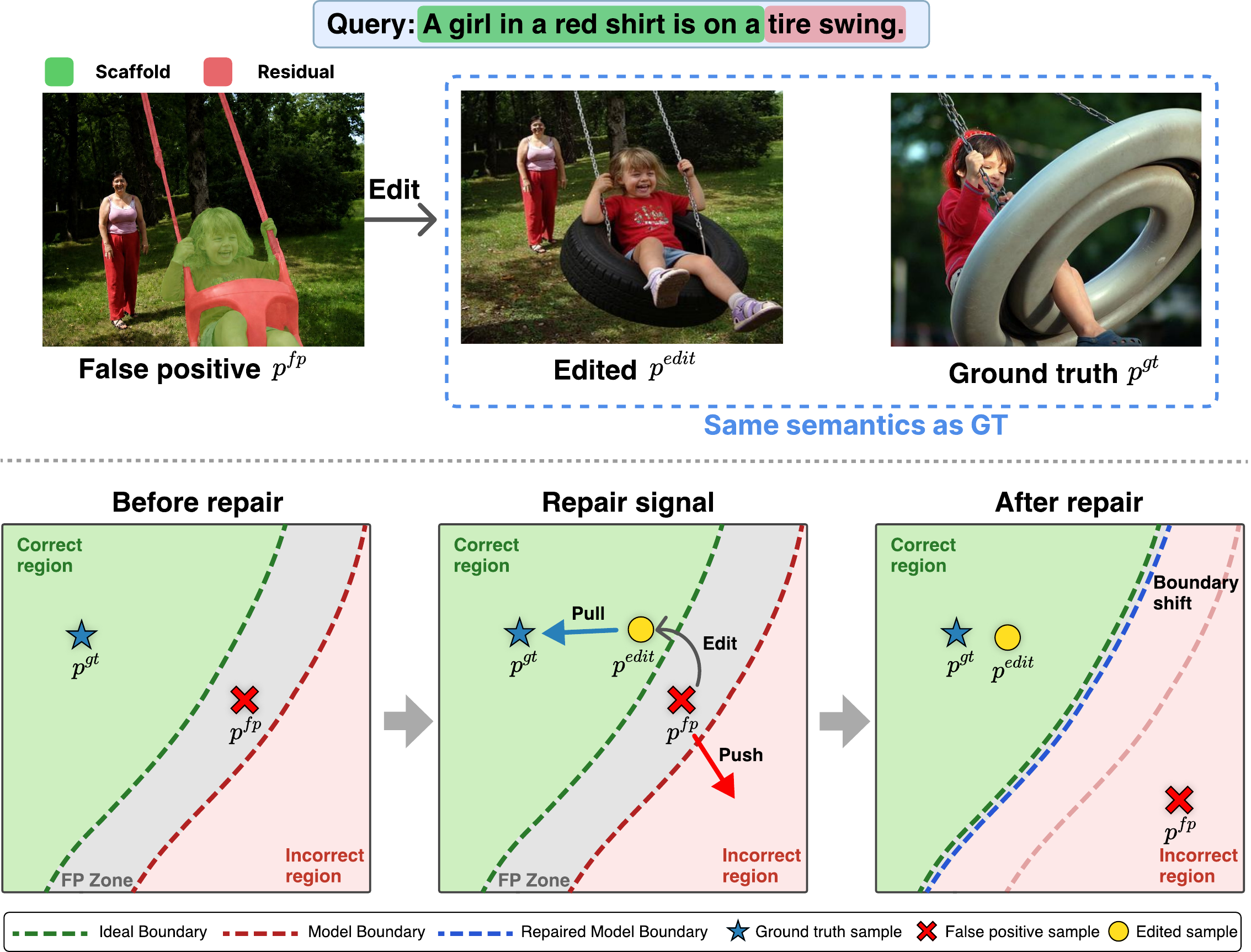}
\vspace{-10pt}
\caption{\textbf{The RePair operator.} \textbf{Top:} For a text query $q$, the retrieval system returns a near-miss false positive $p^{\mathrm{fp}}$ that shares the query's scaffold (green: ``a girl in a red shirt'') but differs on the failure-causing residual (red: wrong swing type). RePair minimally edits only the residual to produce $p^{\mathrm{edit}}$, which carries the same semantics as the ground truth $p^{\mathrm{gt}}$. \textbf{Bottom:} Before repair, $p^{\mathrm{fp}}$ falls in the FP zone between the ideal boundary and the model's learned boundary. The repair signal pulls $p^{\mathrm{edit}}$ toward $p^{\mathrm{gt}}$ and pushes $p^{\mathrm{fp}}$ away, shifting the model boundary to eliminate the FP zone.}
\vspace{-15pt}
\label{fig:teaser}
\end{figure}

Vision-language retrieval has become a standard interface for cross-modal search, with CLIP-style dual encoders serving as the dominant architecture~\cite{radford2021learning,jia2021scaling}. While these models are effective at matching high-level visual and textual concepts, practical retrieval often hinges on localized semantic distinctions---attribute binding, relations, and word order---on which diagnostic benchmarks document persistent failures~\cite{thrush2022winoground,yuksekgonul2023when}. The critical details can be subtle: whether a person is alone or in a group, whether an object is a projector or a skateboard, or who is performing versus receiving an action. Under such demands, the most consequential errors are not obviously unrelated candidates, but top-ranked near misses that satisfy most of the query while missing the detail that makes the match correct (see Fig.~\ref{fig:teaser}, top, for an example). These near misses are especially informative because they reveal precisely where the retriever's semantic understanding breaks down.

Near misses therefore provide a natural handle for refining retrieval models. Standard contrastive fine-tuning relies largely on random in-batch negatives, many of which are too easy to sharpen such fine-grained boundaries~\cite{robinson2021contrastive}. Hard-sample mining moves closer to the boundary by retrieving confusable candidates from the training corpus, and has become a common way to strengthen contrastive supervision~\cite{karpukhin-etal-2020-dense,xiong2021approximate}. However, mining---like reweighting~\cite{robinson2021contrastive,chuang2020debiased}---remains a selection-only mechanism: both families select or reweight existing items rather than altering pair content. Mining can identify what the model should push away, but it cannot construct the corrected neighbor in the same semantic region. As a result, the model receives repulsive supervision from hard negatives, but lacks a matched attractive signal that specifies how the boundary should be repaired.

Synthetic augmentation offers a complementary route~\cite{liu-etal-2024-unleashing-power,liu2025structsynth}: instead of only selecting existing candidates, it can generate hard positives or hard negatives~\cite{cioni2023diffusion,yoon-etal-2024-assessing,alimisis2025advances}. Yet most existing pipelines are not conditioned on the retriever's own failures. They perturb captions, rewrite texts, synthesize images, or compose counterfactuals according to predefined templates or external generators, without asking which semantic factor actually caused the current model to fail~\cite{yuksekgonul2023when,fan2023improving,patel2024tripletclip,tian2023stablerep}. Consequently, the generated samples may be plausible hard cases in general, but not necessarily hard for the model being trained. They may alter irrelevant details while missing the true confusion factor, leading to low hit rates and large synthesis volumes~\cite{chuang2020debiased}.

Our key observation is that a top-ranked false positive is a useful counterfactual scaffold, not merely a negative. It preserves much of the query's semantic structure, explaining why the model retrieved it, while containing a localized failure-causing residual, explaining why it is wrong. This scaffold--residual view suggests a simple operator over a retrieval error: given a query $q$ and its retrieved false positive $p^{\mathrm{fp}}$, minimally correct only the failure-causing residual of $p^{\mathrm{fp}}$ while preserving its shared scaffold. As illustrated in Fig.~\ref{fig:teaser}, because the corrected result $p^{\mathrm{edit}}$ now carries the same semantic content as the ground truth $p^{\mathrm{gt}}$, it forms a hard positive of $p^{\mathrm{gt}}$ in the same modality; meanwhile, $p^{\mathrm{edit}}$ and the unedited $p^{\mathrm{fp}}$ form a hard negative pair that straddles the decision boundary with minimal perceptual difference. Training can therefore pull $p^{\mathrm{edit}}$ toward the query alongside $p^{\mathrm{gt}}$ and push $p^{\mathrm{fp}}$ away, shifting the model boundary to eliminate the FP zone where the failure occurred (Fig.~\ref{fig:teaser}, bottom).

We instantiate this idea as \textbf{RePair}. The name reflects two coupled operations: RePair \textbf{re-pairs} each retrieval error with a minimal counterfactual counterpart, and training on the resulting contrast \textbf{repairs} the local retrieval boundary by pulling the corrected counterpart toward the query alongside the ground truth while pushing the original error away. To make this operation reliable, RePair follows three design principles. \textbf{Validity} ensures that a mined near miss is a genuine retrieval failure rather than annotation noise or a benign alternative match. \textbf{Minimality} ensures that counterfactual editing changes only the failure-causing residual while preserving the shared semantic scaffold. \textbf{Locality} ensures that the original failure and its corrected counterpart are contrasted within the same boundary neighborhood. Operationally, RePair mines top-ranked false positives in both retrieval directions and verifies each with an LLM-based semantic auditor (\textbf{Validity}). It then generates minimal edit instructions and selects the least-disruptive valid edit (\textbf{Minimality}). Finally, it trains the resulting hard pairs with a local contrastive objective that couples pull and push signals within the same boundary neighborhood (\textbf{Locality}).

We evaluate RePair on Flickr30K and COCO30K under a unified CLIP fine-tuning setting. RePair achieves the strongest top-rank (R@1) retrieval performance among controlled augmentation baselines while using substantially fewer synthetic samples. Ablations confirm that failure conditioning, quality control, and paired pull--push supervision drive the gains rather than synthesis volume alone. These results support the central premise of RePair: retrieval failures are not just mistakes to be discarded or negatives to be mined, but compact opportunities for counterfactual boundary repair. In summary, we make the following contributions:

\begin{itemize}[leftmargin=*]
\item We introduce \textbf{RePair}, a failure-conditioned counterfactual training framework that re-pairs a retriever's own false positives with minimal counterfactual counterparts, converting each retrieval error into a hard positive of the ground truth and a grounded hard negative pair for local boundary repair.
\item We instantiate RePair through three design principles---Validity, Minimality, and Locality---using LLM-based semantic auditing, minimal counterfactual editing with post-edit selection, bidirectional hard-pair construction, and local hard-pair contrastive training.
\item We show that RePair improves vision-language retrieval more data-efficiently than error-agnostic augmentation under a unified CLIP fine-tuning setting, with ablations isolating the roles of failure conditioning, quality control, and paired pull--push supervision.
\end{itemize}

%% file: sections/related-work.tex
\section{Related Work}
\label{sec:related_work}

\subsection{Hard-sample Aware Contrastive Vision-Language Learning}

In contrastive vision-language learning~\cite{radford2021learning,jia2021scaling}, the training signal is dominated by \textit{hard negatives}---samples that are semantically distinct yet close in the embedding space---while simply increasing batch size yields diminishing returns~\cite{robinson2021contrastive,kalantidis2020mixing,wu2017sampling}.
Modern strategies such as hard negative mining, filtering, and distillation improve data efficiency and robustness~\cite{faghri2018vse++,radenovic2023filtering,wang2025getting}.
Despite these advances, most existing approaches improve hard signals at the distribution or strategy level, but do not condition on the model's \textit{current retrieval failures} as explicit anchors for failure-conditioned synthesis.

\begin{figure*}[t]
\centering
\includegraphics[width=\textwidth]{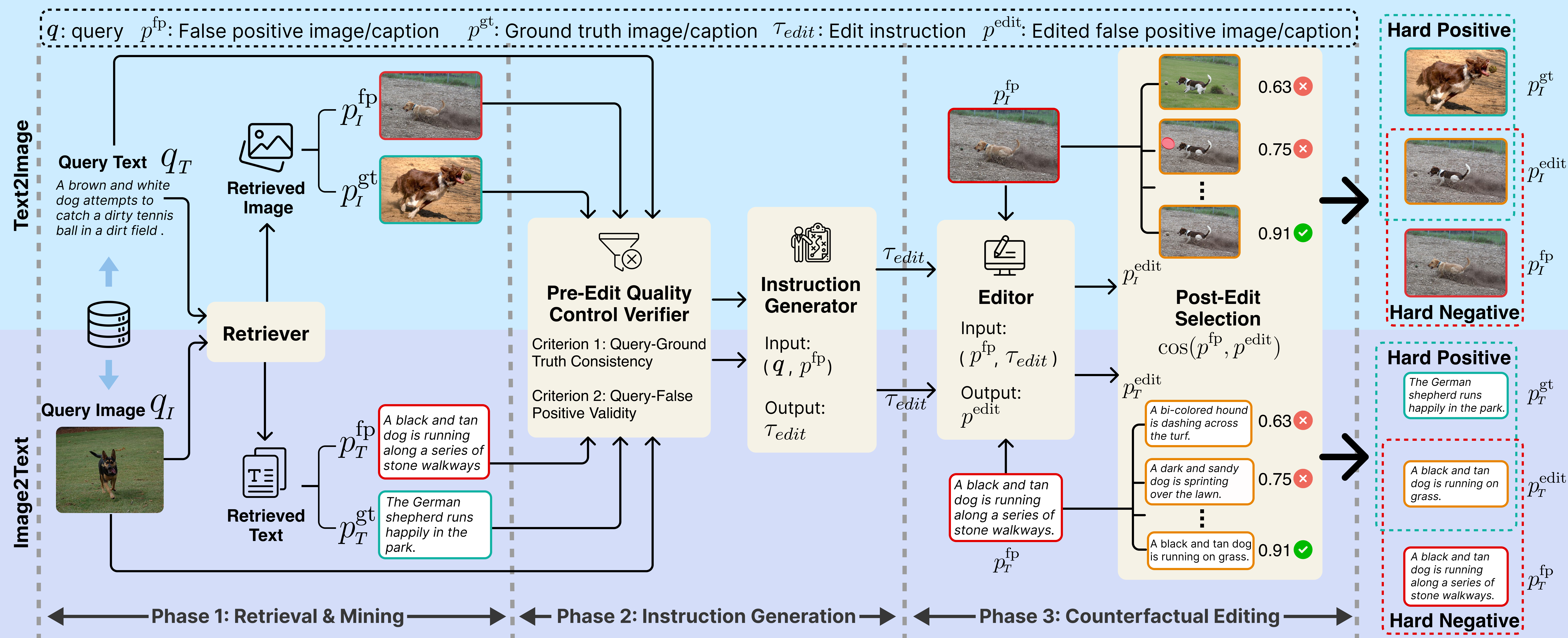}
\caption{RePair pipeline overview.
\textbf{Validity (Mining \& Validation)}: A retriever mines top-$K_{\mathrm{mine}}$
false positives ($p^{\mathrm{fp}}_I$/$p^{\mathrm{fp}}_T$) in both T2I and
I2T directions; a pre-edit verifier filters annotation noise and
benign mismatches, ensuring only genuine failures serve as editing seeds.
\textbf{Minimality (Counterfactual Editing)}: An instruction generator
produces a minimal edit instruction $\tau_{\mathrm{edit}}$ conditioned
on $(q,p^{\mathrm{fp}})$; a modality-specific editor
applies $\tau_{\mathrm{edit}}$ to $p^{\mathrm{fp}}$, and post-edit
selection picks the candidate closest to $p^{\mathrm{fp}}$, yielding
a hard positive $p^{\mathrm{edit}}$ (edited) and a hard negative
$p^{\mathrm{fp}}$ (original). The resulting pairs are organized into
local contrastive grids for training (\textsc{Locality};
see Figure~\ref{fig:block_matrix}).}
\label{fig:pipeline}
\end{figure*}

\subsection{Synthetic Hard Samples for Vision-Language Models}

To address the scarcity of naturally occurring hard samples, recent work leverages generative models to synthesize hard samples across modalities.
On the text side, NegCLIP~\cite{yuksekgonul2023when} and SugarCrepe~\cite{hsieh2023sugarcrepe} generate hard negative captions via compositional perturbation~\cite{ma2023crepe,donmez-etal-2023-hnc,zhang-etal-2024-countercurate}, while LaCLIP~\cite{fan2023improving} produces hard positive captions through paraphrasing.
On the image side, StableRep~\cite{tian2023stablerep} and ALiA~\cite{Dunlap23alia} generate synthetic positive images~\cite{jin2024armada}, while COCO-CF~\cite{le2023coco} and TripletCLIP~\cite{patel2024tripletclip} construct counterfactual or hard negative images~\cite{koohpayegani2025genie,huang2025visual,zhang2025cfvlm}.
However, most methods operate in a single modality and skew toward negatives without a mechanism to balance repulsion from hard negatives with attraction to hard positives~\cite{kamath2024hard}.

%% file: sections/method.tex
\section{Method}
\label{sec:method}

We consider \textbf{bidirectional image--text retrieval} with a dual-encoder architecture. Given a paired dataset $\mathcal{D}=\{(I_i,T_i)\}_{i=1}^N$,
image and text encoders produce $\ell_2$-normalized embeddings, respectively $z_I(I)\!\in\!\mathbb{R}^D$ and $z_T(T)\!\in\!\mathbb{R}^D$, which are instantiated using the CLIP dual-encoder backbone. The scaled logit is $\ell(I,T)=\alpha\, z_I(I)^{\!\top} z_T(T)$, where $\alpha>0$ is a learnable logit scale ($\tau=1/\alpha$ is the temperature).

Our framework, \textbf{RePair}, preserves the standard CLIP contrastive objective but fundamentally changes \emph{how} training pairs are constructed. Building on the scaffold--residual view of retrieval failures (Section~\ref{sec:introduction}), RePair follows three design principles. \textbf{Validity}: we mine the model's top-$K_{\mathrm{mine}}$ false positives in both retrieval directions and verify each as a genuine confusion via LLM-based quality control (Section~\ref{sec:validity}). \textbf{Minimality}: we apply counterfactual editing that changes only the failure-causing residual while preserving the shared scaffold, and select the least-disruptive edit (Section~\ref{sec:minimality}). \textbf{Locality}: we organize the resulting hard pairs into local contrastive grids that confine pull--push supervision to the same boundary neighborhood (Section~\ref{sec:locality}). Figure~\ref{fig:pipeline} illustrates the pipeline.

\subsection{Failure-Driven Mining and Validation (\textsc{Validity})}
\label{sec:validity}

Not all retrieval errors are equally useful for training. High-scoring false positives may arise from annotation noise, where the ground truth is mismatched with the query, or from benign mismatches, where the false positive is actually a valid alternative match. The \textsc{Validity} principle ensures that only genuine model confusions---cases where the retriever is systematically wrong---serve as seeds for counterfactual editing.

We adopt a \textbf{model-in-the-loop} strategy: querying the current retrieval system and treating high-scoring mismatched results as failure anchors. Each top-ranked false positive $p^{\mathrm{fp}}$ preserves much of the query's semantic structure---the shared scaffold that explains its high ranking---while differing in a localized failure-causing residual. This scaffold--residual structure makes $p^{\mathrm{fp}}$ an ideal seed for counterfactual editing (Section~\ref{sec:minimality}).

Given a query $q$, let $\mathcal{P}(q)$ denote all its annotated positives and let $\mathcal{C}(q)$ denote the cross-modal candidate set (all images when $q$ is text, all captions when $q$ is an image). Excluding all annotated positives, we rank the remaining candidates by similarity and keep the top $K_{\mathrm{mine}}$ as the \emph{false-positive pool}:
\begin{equation}
\mathcal{F}(q) = \operatorname*{Top}_{p \in \mathcal{C}(q) \setminus \mathcal{P}(q)}^{\,K_{\mathrm{mine}}} \, \ell(q, p),
\label{eq:fp_pool_image}
\end{equation}
where $\operatorname{Top}^{n}$ returns the $n$ candidates with the highest scores.
From this pool, we retain the top-$d$ most confusable false positives:
\begin{equation}
\{p_i^{\mathrm{fp}}\}_{i=1}^d
= \operatorname*{Top}_{p \in \mathcal{F}(q)}^{\,d} \, \ell(q, p),
\label{eq:hard_fp}
\end{equation}
where $\ell(q,p_1^{\mathrm{fp}}) \ge \cdots \ge \ell(q,p_d^{\mathrm{fp}})$.
The retrieval pool size satisfies $K_{\mathrm{mine}} \ge d$, so that the pool always supplies the $d$ retained false positives. The retention count $d$ is direction-dependent: text editing via an LLM is cheap and deterministic, permitting a larger $d$, whereas image editing via a diffusion model is expensive and stochastic, favoring a smaller $d$ combined with multiple candidate generations (Section~\ref{sec:minimality}) for diversity.

\paragraph{Pre-Edit Quality Control.}
We prompt an LLM to evaluate two criteria in a single pass (see Appendix~\ref{prompt:qc}):
\textbf{Criterion~1} verifies query--ground-truth consistency; \textbf{Criterion~2} checks whether the false positive constitutes a valid alternative match.
We retain a triplet only when the query--ground-truth pair is confirmed valid and the false positive is \emph{not} a valid alternative match, ensuring that only genuine model confusions are forwarded to the editing stage.

\subsection{Minimal Counterfactual Editing (\textsc{Minimality})}
\label{sec:minimality}

The \textsc{Minimality} principle requires that counterfactual editing changes only the failure-causing residual while preserving the shared scaffold between the false positive and the query. This ensures that the resulting hard positive $p^{\mathrm{edit}}$ is perceptually near-identical to the original false positive $p^{\mathrm{fp}}$, differing in exactly the semantic attribute that caused the retrieval error.

\paragraph{Edit-Instruction Generation.}
For each verified failure, an instruction generator $G(\cdot)$ produces a counterfactual editing instruction $\tau_{\mathrm{edit}} = G(q, p^{\mathrm{fp}})$, specifying how to minimally modify $p^{\mathrm{fp}}$ to match query $q$.

\paragraph{Counterfactual Editing and Post-Edit Selection.}
A modality-specific editor $E(\cdot)$---an LLM for text, a diffusion model for images---generates $M$ edited candidates $\{p^{\mathrm{edit}}_m\}_{m=1}^M = \{E(p^{\mathrm{fp}}, \tau_{\mathrm{edit}}; \xi_m)\}_{m=1}^M$ by varying stochastic factors~$\xi_m$.
From the $M$ candidates, we select the final hard example by the \textbf{minimal-change} principle:
\begin{equation}
p^{\mathrm{edit}}
= \operatorname*{arg\,max}_{m \in \{1,\dots,M\}}
\operatorname{cos}\!\bigl(
z(p^{\mathrm{fp}}),\; z(p^{\mathrm{edit}}_m)
\bigr).
\label{eq:select_hp}
\end{equation}

The edit instruction encourages query alignment, while the minimal-change selection encourages scaffold preservation. Together they balance semantic correction with structural fidelity. We term this selection step \textbf{post-edit quality control (PQC)}.

Because the edit instruction targets only the failure-causing residual, the edited counterpart $p^{\mathrm{edit}}$ shares the same scaffold as the original $p^{\mathrm{fp}}$. The corrected $p^{\mathrm{edit}}$ carries the same semantic content as the ground truth $p^{\mathrm{gt}}$, forming a hard positive of $p^{\mathrm{gt}}$ in the same modality; at the same time, $p^{\mathrm{edit}}$ and $p^{\mathrm{fp}}$ form a hard negative pair whose hardness is grounded in their minimal residual contrast.
The synthesis pipeline outputs a grounded tuple $(q, p^{\mathrm{gt}}, p^{\mathrm{fp}}, p^{\mathrm{edit}})$.
The full synthesis procedure is in Algorithm~\ref{alg:synthesis} (Appendix~\ref{app:algorithms}).

\subsection{Local Hard-Pair Contrastive Training (\textsc{Locality})}
\label{sec:locality}

The synthesis stages above produce hard pairs that are semantically proximate by construction---each pair shares a scaffold and differs only in the failure-causing residual. However, random batching does not guarantee that such a pair appears in the same minibatch, diluting the paired pull--push signal. The \textsc{Locality} principle addresses this by organizing synthesized examples into structured local neighborhoods where the hard positive and its counterfactual hard negative are always contrasted together.

\paragraph{Bidirectional Hard-Pair Construction.}
The synthesis pipeline of Sections~\ref{sec:validity}--\ref{sec:minimality} is applied in both retrieval directions. For a T2I false positive (image $I^{\mathrm{fp}}$ retrieved for text query $T$), the corrected result $I^{+}$ forms a hard positive of the ground-truth image in the same modality, while $I^{+}$ and $I^{\mathrm{fp}}$ form a hard negative pair. Symmetrically, for an I2T false positive (caption $T^{\mathrm{fp}}$ retrieved for image query $I$), the corrected $T^{+}$ forms a hard positive of the ground-truth caption, with $T^{+}$ and $T^{\mathrm{fp}}$ forming a hard negative pair. This bidirectional process produces complementary hard pairs across both modalities.

\begin{figure}[t]
\centering
\includegraphics[width=\columnwidth]{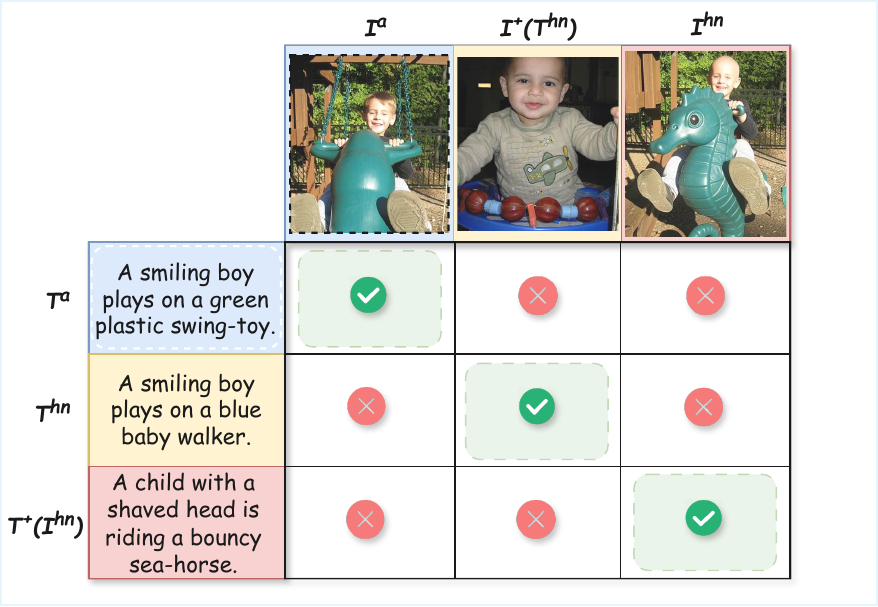}
\caption{The $3\times 3$ local confusion grid. Diagonal entries are positive pairs; off-diagonal entries serve as hard negatives.}
\label{fig:block_matrix}
\end{figure}

\paragraph{Grid Construction.}
Starting from a ground-truth anchor pair $(I^{\mathrm{a}},T^{\mathrm{a}})$, we incorporate the mined false-positive image $I^{\mathrm{hn}}$\footnote{We write $I^{\mathrm{hn}}$/$T^{\mathrm{hn}}$ instead of $I^{\mathrm{fp}}$/$T^{\mathrm{fp}}$ to mark the shift from synthesis role (edit seed) to training role (hard negative for the grid anchor). When obtained via editing, $I^{+}(T^{\mathrm{hn}})$ and $T^{+}(I^{\mathrm{hn}})$ are the post-edit-selected outputs of the synthesis records in which $T^{\mathrm{hn}}$ and $I^{\mathrm{hn}}$ themselves served as queries (cf.\ Algorithm~\ref{alg:training}).} for $T^{\mathrm{a}}$ (T2I failure) and the false-positive text $T^{\mathrm{hn}}$ for $I^{\mathrm{a}}$ (I2T failure). Each false positive is associated with a semantically matching counterpart in its cross-modal partner: $I^{+}(T^{\mathrm{hn}})$ denotes the image that semantically matches $T^{\mathrm{hn}}$, and $T^{+}(I^{\mathrm{hn}})$ denotes the caption that semantically matches $I^{\mathrm{hn}}$---each obtained either from the dataset's own annotations or via counterfactual editing. The three resulting pairs form the grid; only diagonal entries are treated as positives, while the six off-diagonal pairs serve as hard negatives. The grid is
\begin{equation}
\begin{aligned}
\mathcal{I} &= [I^{\mathrm{a}},\, I^{+}(T^{\mathrm{hn}}),\, I^{\mathrm{hn}}], \\
\mathcal{T} &= [T^{\mathrm{a}},\, T^{\mathrm{hn}},\, T^{+}(I^{\mathrm{hn}})].
\end{aligned}
\label{eq:block_structure}
\end{equation}
\paragraph{Training Strategy.}
Directly mixing synthesized hard examples into the global batch risks degrading feature-space uniformity~\cite{wang2020understanding}. The decoupled grid confines hard-pair interactions to a local neighborhood, preserving global uniformity while sharpening local alignment.
We therefore design a \textbf{dual-objective framework}: the global contrastive loss $\mathcal{L}_{\mathrm{global}}$, the standard symmetric InfoNCE loss, operates exclusively on original anchor pairs, while synthesized hard examples participate only in the local grid objective.

\paragraph{Local $3\times 3$ Grid Loss.}
For each grid $b \in \{1, \dots, B\}$, let $\mathbf{S}^{(b)} \in \mathbb{R}^{3\times 3}$ be the pairwise logit matrix within that grid. Let $\mathcal{Q}$ denote the set of diagonal positive pairs, which is the same for every grid.
The local loss forces the model to discriminate anchor and hard positives from hard negatives within the grid:
\begin{equation}
\mathcal{L}_{\mathrm{grid}}
= \frac{1}{B} \sum_{b=1}^B \frac{1}{2|\mathcal{Q}|}\sum_{(i,j)\in\mathcal{Q}} \mathcal{J}^{(b)}_{ij},
\label{eq:block_loss}
\end{equation}
where
\begin{equation*}
\mathcal{J}^{(b)}_{ij} = -\log\frac{e^{S^{(b)}_{ij}}}{\sum_{k=1}^3 e^{S^{(b)}_{ik}}}
-\log\frac{e^{S^{(b)}_{ij}}}{\sum_{k=1}^3 e^{S^{(b)}_{kj}}}.
\end{equation*}

The final objective is a weighted sum:
\begin{equation}
\mathcal{L} = \mathcal{L}_{\mathrm{global}} + \lambda \mathcal{L}_{\mathrm{grid}},
\label{eq:final_loss}
\end{equation}
where $\lambda \ge 0$ balances the global and local supervision.
The full training procedure is in Algorithm~\ref{alg:training} (Appendix~\ref{app:algorithms}).

%% file: sections/experiments.tex
\section{Experiments}
\begin{table*}[t]
    \centering
    \small
    \setlength{\tabcolsep}{4pt}
    \renewcommand{\arraystretch}{1.08}
    \begin{tabular}{lc cccc cc cc cc cc}
    \toprule
    \multirow{2}{*}{Method} &
    \multirow{2}{*}{Samples (k)} &
    \multicolumn{4}{c}{Aug. Type} &
    \multicolumn{4}{c}{Flickr30K} &
    \multicolumn{4}{c}{COCO30K} \\
    \cmidrule(lr){3-6} \cmidrule(lr){7-10} \cmidrule(lr){11-14}
    & & $I^{+}$ & $I^{-}$ & $T^{+}$ & $T^{-}$
    & \multicolumn{2}{c}{I2T} & \multicolumn{2}{c}{T2I}
    & \multicolumn{2}{c}{I2T} & \multicolumn{2}{c}{T2I} \\
    & & & & & & R@1 & R@5 & R@1 & R@5 & R@1 & R@5 & R@1 & R@5 \\
    \midrule
    Vanilla
    & -- & & & & &
    87.87 & 98.22 &
    74.60 & 92.87 &
    60.54 & 83.86 &
    44.15 & 71.97 \\
    
    NegCLIP
    & 205 & & & & \cmark &
    88.07 & 98.13 &
    74.38 & 92.78 &
    61.06 & 84.00 &
    44.37 & 72.29 \\
    
    SugarCrepe
    & 145 & & & & \cmark &
    88.95 & \second{98.62} &
    74.64 & 92.98 &
    61.92 & 85.02 &
    45.09 & 72.94 \\
    
    LaCLIP
    & 145 & & & \cmark & &
    86.69 & 97.83 &
    74.46 & 92.60 &
    60.96 & 84.30 &
    44.99 & 72.70 \\
    
    SimCLR
    & 145 & \cmark & & & &
    88.26 & 98.32 &
    74.30 & 92.78 &
    61.34 & 84.14 &
    45.13 & 72.77 \\
    
    StableRep
    & 145 & \cmark & & & &
    87.67 & 98.22 &
    74.64 & 92.72 &
    61.76 & 84.98 &
    44.61 & 72.95 \\
    
    ALiA
    & 145 & \cmark & & & &
    87.67 & 98.22 &
    74.36 & 92.88 &
    61.62 & 84.40 &
    44.93 & 72.87 \\
    
    COCO-CF
    & 145 & & \cmark & & \cmark &
    87.97 & 97.44 &
    74.64 & 92.98 &
    60.32 & 84.92 &
    44.55 & 72.61 \\
    
    TripletCLIP
    & 145 & & \cmark & & \cmark &
    85.11 & 97.44 &
    73.47 & 92.13 &
    56.06 & 84.40 &
    43.20 & 71.23 \\
    \midrule
    \multicolumn{14}{l}{\textit{Combined Baselines}} \\
    La + SC
    & 290 & & & \cmark & \cmark &
    88.07 & 97.34 &
    74.24 & 92.62 &
    60.10 & 84.96 &
    44.57 & 72.63 \\

    CF + AL + La
    & 435 & \cmark & \cmark & \cmark & \cmark &
    86.88 & 97.63 &
    74.14 & 92.84 &
    59.74 & 83.98 &
    44.56 & 72.59 \\

    \midrule
    
    RePair-I2T
    & 47 & & & \cmark & \cmark &
    88.75 & 98.12 &
    \second{76.07} & \best{93.73} &
    \second{62.06} & 84.84 &
    45.69 & 73.31 \\
    
    RePair-T2I
    & 60 & \cmark & \cmark &  &  &
    \second{89.44} & 98.32 &
    75.93 & 93.47 &
    61.78 & \second{85.24} &
    \second{45.87} & \second{73.53} \\
    
    RePair (Full)
    & 107 & \cmark & \cmark & \cmark & \cmark &
    \best{90.13} & \best{98.71} &
    \best{76.35} & \second{93.67} &
    \best{62.64} & \best{85.72} &
    \best{46.11} & \best{73.59} \\
    
    \bottomrule
    \end{tabular}
    \caption{Retrieval performance on Image-to-Text (I2T) and Text-to-Image (T2I) on Flickr30K and COCO30K. The best and second-best results are marked in \textbf{bold} and \underline{underlined}, respectively. Augmentation modality and polarity are marked by $I^{+}$/$I^{-}$ (image hard positive/negative) and $T^{+}$/$T^{-}$ (text hard positive/negative).}
    \label{tab:merged_results}
\end{table*}

\subsection{Experimental Setup}

\paragraph{Datasets.}
We evaluate on \textbf{Flickr30K}~\cite{Plummer2015Flickr30k} and \textbf{MS-COCO}~\cite{Lin2014COCO}. To align the training scale with Flickr30K (29k training images), we construct \textbf{COCO30K} by uniformly sampling from the COCO training split while maintaining the original image--caption pairing (see Appendix~\ref{app:datasets}). For evaluation, we strictly follow the standard Karpathy split~\cite{karpathy2015deep} for both datasets.

\paragraph{Evaluation Protocol.}
We perform full-rank retrieval on the test set for both Image-to-Text (I2T) and Text-to-Image (T2I) directions. Detailed protocol settings are provided in Appendix~\ref{app:eval}.

\paragraph{Baselines.}
We compare our approach against three categories of strong baselines:
(a) \textbf{Text-side synthesis/rewrites}: NegCLIP~\cite{yuksekgonul2023when}, SugarCrepe~\cite{hsieh2023sugarcrepe}, and LaCLIP~\cite{fan2023improving};
(b) \textbf{Image-side augmentation/synthesis}: SimCLR~\cite{chen2020simple}, StableRep~\cite{tian2023stablerep}, and ALiA~\cite{Dunlap23alia};
(c) \textbf{Pair-level / pipeline}: COCO-CF~\cite{le2023coco} and TripletCLIP~\cite{patel2024tripletclip}.
We also evaluate representative composed baselines (e.g., La\,+\,SC; CF\,+\,AL\,+\,La, with La\,=\,LaCLIP, SC\,=\,SugarCrepe, CF\,=\,COCO-CF, AL\,=\,ALiA) to analyze complementarity. Implementation recipes for all baselines are detailed in Appendix~\ref{app:baselines}.

\paragraph{Metrics.}
We report Recall@K ($K \in \{1, 5, 10\}$) and Mean Reciprocal Rank (MRR) for both I2T and T2I. Full results including R@10 and MRR are provided in Appendix~\ref{app:full_results}.

\paragraph{Implementation Details.}
We use the \textsf{CLIP ViT-B/32}~\cite{radford2021learning} backbone for all main experiments unless otherwise noted. For the synthesis pipeline, we employ \textsf{Gemini 2.5 Flash}~\cite{comanici2025gemini} for quality control, instruction generation, and I2T caption editing, while \textsf{Flux2-9B}~\cite{flux-2-2025} is used for image editing. We set $K_{\mathrm{mine}}{=}100$ for the retrieval pool (Eq.~\ref{eq:fp_pool_image}) and use asymmetric retention counts (Eq.~\ref{eq:hard_fp}): $d_{\text{I2T}}{=}5$ (retaining the 5 highest-scoring false-positive captions) and $d_{\text{T2I}}{=}1$ (retaining the single top-ranked false-positive image). The asymmetric $d$ reflects the modality cost structure discussed in Section~\ref{sec:validity}. This yields approximately 47k I2T and 60k T2I synthesis instances (totaling 107k). For post-edit selection (PQC, Eq.~\ref{eq:select_hp}), we generate $M{=}3$ candidates and retain the one closest to $p^{\mathrm{fp}}$ in embedding space. We set $\lambda=0.5$ for the $3 \times 3$ grid loss. Additional hyperparameters and training details are provided in Appendix~\ref{app:hyperparams}.

\subsection{Main Results}

Table~\ref{tab:merged_results} presents the retrieval performance comparison on Flickr30K and COCO30K. We evaluate three variants of our framework: \textbf{RePair-I2T} (applying synthesis only to Image-to-Text failures), \textbf{RePair-T2I} (applying synthesis only to Text-to-Image failures), and \textbf{RePair (Full)} (the complete bidirectional framework). Our approach demonstrates three key advantages over existing augmentation strategies.

\begin{itemize}[leftmargin=*]
    \item \textbf{Data-efficient superiority.} With only \textbf{107K} synthetic samples---26\%--75\% fewer than the 145K--435K used by baselines---RePair~(Full) achieves the best R@1 across both datasets and both retrieval directions (e.g., Flickr30K I2T: 90.13 vs.\ SugarCrepe's 88.95). These gains confirm that failure-conditioned synthesis targeting the model's error space yields higher-quality supervision than scaling volume. These gains are most pronounced at R@1, the regime most sensitive to the near-miss confusions that failure conditioning targets.

    \item \textbf{Bidirectional complementarity.} Each single-direction variant improves \emph{both} retrieval directions, lifting cross-direction R@1 by more than a point in every case---indicating that denser supervision on one confusion boundary indirectly sharpens the shared embedding space. Combining both directions lifts R@1 in all four dataset--direction settings; the two directions therefore supply complementary supervision.

    \item \textbf{Precision over volume.} Composite baselines using 290K--435K samples fail to improve consistently over Vanilla and can degrade below it (e.g., CF\,+\,AL\,+\,La on Flickr30K I2T R@1: 86.88 vs.\ 87.87), demonstrating that error-agnostic synthesis introduces noise that pollutes contrastive gradients. In contrast, RePair's failure-conditioned refinement avoids such negative transfer. A sample-size analysis (Appendix~\ref{app:sample_size}) and generation-cost comparison (Appendix~\ref{app:cost_comparison}) further quantify RePair's efficiency advantage.
\end{itemize}

\subsection{Ablation and Analysis}
\label{sec:ablation}

Table~\ref{tab:ablation_r1} reports a component-wise ablation on \textbf{COCO30K}. We fix the synthesis budget, training schedule, and all hyperparameters, toggling one design choice at a time.

\begin{itemize}[leftmargin=*]
    \item \textbf{(i) Failure conditioning (\textsc{Validity}) is the dominant factor.} Replacing failure-mined seeds with error-agnostic sampling (\textit{w/o Failure}) causes the largest drop (I2T: $-2.10$, T2I: $-2.42$), identifying the \textsc{Validity} principle---targeting the model's current error regions rather than generic samples---as the primary driver of RePair's gains. A characterization of failure types is provided in Appendix~\ref{app:failure_char}. Conversely, using the mined false positives directly as hard negatives without editing (\textit{w/o Repair}) yields only marginal gains over Vanilla (Table~\ref{tab:ablation_r1}); the repair step---the corrected hard positive and its minimal contrast against the FP---contributes the bulk of the improvement, confirming that mining and repairing are both necessary.

    \item \textbf{(ii) Seed validation (\textsc{Validity}) and minimal-change selection (\textsc{Minimality}) are both necessary.} Removing seed-level filtering (\textit{w/o QC}) or post-edit minimal-change selection (\textit{w/o PQC}) consistently hurts both directions (Table~\ref{tab:ablation_r1}), with larger T2I degradations suggesting image-side edits are more prone to semantic drift. QC reliability evaluation is provided in Appendix~\ref{app:qc_eval}.

    \item \textbf{(iii) Local paired supervision (\textsc{Locality}) is essential.} Discarding the local grid (\textit{Full Matrix}) produces the largest single-component drop after failure conditioning (Table~\ref{tab:ablation_r1})---evidence that mixing synthetic items into the global batch dilutes paired pull--push supervision. Retaining the grid but removing hard-positive supervision (\textit{Hard Neg Only}) further shows that push-only training is insufficient---the model requires the complementary pull signal from counterfactual hard positives.
\end{itemize}

\paragraph{Sensitivity to external components.}
Swapping the image editor introduces notable variance---InstructPix2Pix~\cite{brooks2023instructpix2pix} degrades markedly while Gemini~2.5 Flash Image yields slight gains---indicating that RePair is plug-and-play but its ceiling scales with editor fidelity. The characteristic failure mode of image editing is scaffold drift, found only in T2I; better editors directly reduce this failure mode, raising RePair's ceiling (Appendices~\ref{app:edit_failure_modes} and~\ref{app:editor_sensitivity}). Replacing the LLM with GPT-4o~\cite{openai2024gpt4o} degrades R@1 modestly (Table~\ref{tab:ablation_r1})---well under half the drop caused by removing failure mining entirely---confirming the core gain comes from failure-conditioned synthesis, not specific LLM capability. On ViT-L/14, RePair improves both COCO30K directions (Table~\ref{tab:ablation_r1}); the gains persist on this stronger backbone, though narrower margins are expected as the baseline leaves fewer near-miss errors to repair. Multi-round re-mining yields additional but diminishing gains, suggesting one round captures most informative failures (Appendix~\ref{app:multi_round}).

\begin{table}[t]
    \centering
    \resizebox{\columnwidth}{!}{%
    \begin{tabular}{llcc}
    \toprule
    Category & Ablation setting & $\mathrm{R@1}_{\mathrm{I2T}}$ & $\mathrm{R@1}_{\mathrm{T2I}}$ \\
    \midrule
    RePair & - & 62.64 & 46.11 \\
    \midrule
    Failure Cond.\ (\textsc{Val.})
    & w/o Failure  & $60.54_{\text{\scriptsize \textcolor{BrickRed}{-2.10}}}$ & $43.69_{\text{\scriptsize \textcolor{BrickRed}{-2.42}}}$ \\
    Repair Op.\ (\textsc{Min.}\,$+$\,\textsc{Loc.})
    & w/o Repair (FP-as-HN) & $60.83_{\text{\scriptsize \textcolor{BrickRed}{-1.81}}}$ & $44.20_{\text{\scriptsize \textcolor{BrickRed}{-1.91}}}$ \\
    \midrule
    Seed Valid.\ (\textsc{Val.})
    & w/o QC  & $61.98_{\text{\scriptsize \textcolor{BrickRed}{-0.66}}}$ & $44.87_{\text{\scriptsize \textcolor{BrickRed}{-1.24}}}$ \\
    Min.-Change (\textsc{Min.})
    & w/o PQC & $62.12_{\text{\scriptsize \textcolor{BrickRed}{-0.52}}}$ & $45.33_{\text{\scriptsize \textcolor{BrickRed}{-0.78}}}$ \\
    \midrule
    \multirow{2}{*}{Local Training (\textsc{Loc.})}
    & Full Matrix  & $61.42_{\text{\scriptsize \textcolor{BrickRed}{-1.22}}}$ & $44.36_{\text{\scriptsize \textcolor{BrickRed}{-1.75}}}$ \\
    & Hard Neg Only  & $61.64_{\text{\scriptsize \textcolor{BrickRed}{-1.00}}}$ & $44.92_{\text{\scriptsize \textcolor{BrickRed}{-1.19}}}$ \\
    \midrule
    \multirow{3}{*}{Image Editing Model}
    & Flux2-4B  & $61.82_{\text{\scriptsize \textcolor{BrickRed}{-0.82}}}$ & $45.01_{\text{\scriptsize \textcolor{BrickRed}{-1.10}}}$ \\
    & InstructPix2Pix & $61.07_{\text{\scriptsize \textcolor{BrickRed}{-1.57}}}$ & $43.86_{\text{\scriptsize \textcolor{BrickRed}{-2.25}}}$ \\
    & Gemini 2.5 Flash Image  & $62.86_{\text{\scriptsize \textcolor{ForestGreen}{+0.22}}}$ & $46.35_{\text{\scriptsize \textcolor{ForestGreen}{+0.24}}}$ \\
    \midrule
    LLM
    & GPT-4o  & $61.98_{\text{\scriptsize \textcolor{BrickRed}{-0.66}}}$ & $44.73_{\text{\scriptsize \textcolor{BrickRed}{-1.38}}}$ \\
    \midrule
    \multirow{2}{*}{Backbone (ViT-L/14)}
    & Vanilla & 72.00 & 55.19 \\
    & RePair  & 73.48 & 55.94 \\
    \bottomrule
    \end{tabular}%
    }
    \caption{Ablation and sensitivity analysis on COCO30K. Top: core design ablations (one design choice toggled at a time), each defined where it is discussed in Section~\ref{sec:ablation}. Bottom: external component sensitivity. Subscripts show $\Delta$R@1 vs.\ the full method.}
    \vspace{-5pt}
    \label{tab:ablation_r1}
\end{table}

\paragraph{Post-edit quality.}
The ablations above test these components through downstream accuracy, not through whether the retained pairs are correct. We verify 800 retained post-edit records by hand---200 per dataset $\times$ direction split---against five binary criteria (Appendix~\ref{app:post_edit_audit}): over \textbf{92\% form valid hard contrasts}, with near-perfect FP validity. Consistent with the editor sensitivity above, quality is higher on the I2T side; directional breakdowns and failure-mode classification are in Appendix~\ref{app:edit_failure_modes}.

\begin{figure*}[t]
\centering
\includegraphics[width=0.85\textwidth]{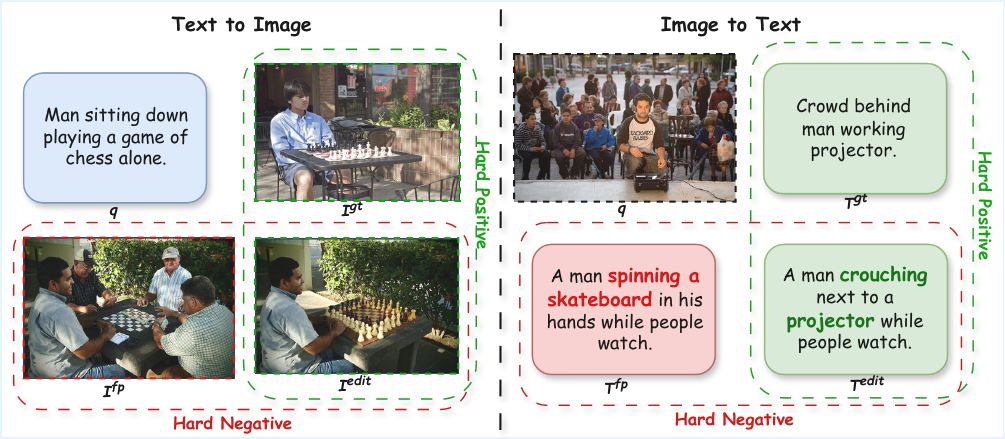}
\caption{Case study of RePair's synthesis on T2I (left) and I2T (right). In T2I, the query $q$ (\textit{``playing chess alone''}) retrieves a near-miss image $I^{\mathrm{fp}}$ depicting multiple players; RePair edits only the failure-causing residual to produce $I^{\mathrm{edit}}$, which matches the query and serves as a hard positive alongside $I^{\mathrm{gt}}$, while the unedited $I^{\mathrm{fp}}$ becomes the paired hard negative. Symmetrically in I2T, a false-positive caption $T^{\mathrm{fp}}$ sharing the query's scaffold (\textit{``people watch''}) but with a mismatched residual (\textit{``skateboard''}) is corrected to a faithful description $T^{\mathrm{edit}}$. $I^{\mathrm{gt}}$ and $T^{\mathrm{gt}}$  denote the ground truth.}
\vspace{-8pt}
\label{fig:case_study}
\end{figure*}

\subsection{Generalization: Scale and Compositionality}
\label{sec:generalization}

\paragraph{Training scale.}
When the training corpus grows $3.9\times$ from COCO30K (29K images) to full MS-COCO (113K), RePair still improves over Vanilla on all four metrics under the same unified setting (e.g., T2I R@1 $+0.93$, R@5 $+0.80$). The gains are consistent but smaller in magnitude, as on ViT-L/14 (Table~\ref{tab:ablation_r1}): scaling the corpus and repairing failures address overlapping near-miss errors. Notably, the improvements persist, indicating that failure-conditioned repair captures error patterns that additional data alone does not fully resolve.

\paragraph{Compositional benchmarks.}
Table~\ref{tab:aro} evaluates the COCO30K-finetuned CLIP ViT-B/32 checkpoints on ARO~\cite{yuksekgonul2023when} under the unified setting of Table~\ref{tab:merged_results}. ARO probes attribute binding, relation reasoning, and word-order sensitivity---the localized distinctions that motivate RePair (Section~\ref{sec:introduction}). RePair improves over Vanilla on all four metrics and ranks first among all ten systems on three---without templated compositional hard negatives. Notably, methods that explicitly construct compositional hard negatives (NegCLIP, SugarCrepe) fall below Vanilla on VG-Attribution, which RePair improves, while the only methods above RePair on VG-Relation (ALiA, SimCLR) remain at Vanilla-level VG-Attribution ($62.4$--$62.5$ vs.\ $64.8$). The relative margins are consistent with the mined failure distribution: relation errors are among the less frequent categories (Appendix~\ref{app:failure_char}).

\begin{table}[t!]
    \centering
    \resizebox{\columnwidth}{!}{%
    \begin{tabular}{lcccc}
    \toprule
    \multirow{2}{*}{Method} & \multicolumn{2}{c}{Accuracy} & \multicolumn{2}{c}{P@1} \\
    \cmidrule(lr){2-3} \cmidrule(lr){4-5}
    & VG-Attr & VG-Rel & COCO-Ord & Flk-Ord \\
    \midrule
    Vanilla       & 62.6 & 51.6 & 25.4 & 31.1 \\
    NegCLIP       & 62.0 & 49.3 & 26.7 & 31.5 \\
    SugarCrepe    & 62.1 & 51.3 & 26.6 & 32.1 \\
    LaCLIP        & \second{62.8} & 49.7 & 27.1 & 32.0 \\
    SimCLR        & 62.5 & \second{53.7} & 27.0 & \second{32.8} \\
    StableRep     & 62.3 & 52.8 & \second{27.4} & 32.4 \\
    ALiA          & 62.4 & \best{54.2} & 27.1 & 32.3 \\
    COCO-CF       & 61.6 & 48.7 & 27.1 & 32.1 \\
    TripletCLIP   & 61.7 & 50.1 & 26.5 & 32.7 \\
    \midrule
    RePair (Full) & \best{64.8} & 53.0 & \best{27.5} & \best{33.2} \\
    \bottomrule
    \end{tabular}%
    }
    \vspace{-6pt}
    \caption{Compositional evaluation on ARO~\cite{yuksekgonul2023when}. All values in \%; best and second-best in \textbf{bold} and \underline{underlined}.}
    \label{tab:aro}
\end{table}

\subsection{Case Study}
Figure~\ref{fig:case_study} illustrates RePair's counterfactual synthesis on representative Flickr30K failures. In the T2I case, a query about \textit{``playing chess alone''} retrieves a near-miss image whose scaffold (outdoor scene, chess board, seated person) matches the query but whose residual (multiple players instead of solitary) causes the failure; RePair edits only this residual ($\text{multi-person} \rightarrow \text{solitary}$), producing $I^{\mathrm{edit}}$ that matches the query, with the unedited $I^{\mathrm{fp}}$ as the paired hard negative. Symmetrically, in the I2T case a false-positive caption preserves the scaffold (\textit{``people watch''}) but contains a mismatched residual (\textit{``skateboard''} instead of \textit{``projector''}); RePair corrects only the residual while keeping the scaffold intact. In both directions, each mined failure produces a tightly controlled pair that concentrates the contrastive gradient on the model's actual confusion boundary.

%% file: sections/conclusion.tex
\section{Conclusion}
\label{sec:conclusion}
We presented \textbf{RePair}, a failure-conditioned counterfactual training framework grounded in the scaffold--residual view of retrieval failures. By observing that a top-ranked false positive preserves most of the query's semantic structure (the scaffold) while differing in a localized failure-causing residual, RePair converts each retrieval error into a hard positive and grounded hard negative through minimal counterfactual editing. Three principles---\textbf{Validity}, \textbf{Minimality}, and \textbf{Locality}---govern seed selection, editing, and training. Experiments on Flickr30K and COCO30K demonstrate that RePair consistently outperforms error-agnostic baselines, particularly at R@1, while using only 107k synthetic samples (26\%--75\% fewer than existing methods), confirming that failure-conditioned counterfactual repair is more data-efficient than brute-force scaling.

%% file: sections/limitations.tex
RePair has several limitations.
First, the synthesis pipeline relies on external editors (an LLM for text, a diffusion model for images) and an LLM-based quality control module. The framework is modular and editor-agnostic---any component can be swapped without architectural changes---so it naturally benefits from advances in the underlying models, but its ceiling is set by their current fidelity.
Second, we mine and repair failures from a single retrieval snapshot; iterative re-mining over successive checkpoints, evaluation on web-scale noisy corpora, and cross-dataset transfer are natural extensions.
Third, the current framework is evaluated on dual-encoder CLIP models; its applicability to cross-attention or fusion-based retrieval architectures remains unexplored.

%% file: sections/ethics-statement.tex
This work focuses on improving image--text retrieval through counterfactual data augmentation. We discuss the ethical aspects of our research below.

\paragraph{Potential Risks and Misuse.}
Our method synthesizes hard negative pairs by editing existing captions and images. While the edits are designed for retrieval training, the underlying text and image editing techniques could be misused to generate misleading content. However, our pipeline operates on well-established public benchmarks (MS-COCO and Flickr30K) and produces training data rather than user-facing content, which limits the scope of potential misuse.

\paragraph{Data and Bias.}
We use publicly available datasets (MS-COCO and Flickr30K) that are widely adopted in the research community. These datasets may contain societal biases present in their source images and annotations. Our method does not address or amplify such biases, as the counterfactual edits target fine-grained semantic distinctions (e.g., attributes, relations, actions) rather than sensitive demographic attributes.

\paragraph{Environmental Impact.}
Our approach is computationally efficient---it does not require training large generative models from scratch but instead leverages existing pre-trained models for data synthesis. We fine-tune the retrieval model on standard hardware with moderate computational cost.

\paragraph{Use of AI Assistants.}
We use a large language model (LLM) as part of our synthesis pipeline for caption editing and quality filtering. In preparing the manuscript, we additionally used AI assistants for language editing and proofreading. All research ideas, experimental design, analyses, and claims are the authors' own.

%% file: sections/acknowledgments.tex
This work was sponsored by the CCF-Tencent Rhino-Bird Open Research Fund
(No.~CCF-Tencent RAGR20250119), and was also supported by the Guangdong Basic
and Applied Basic Research Foundation (No.~2025A1515010304), the Guangdong
Province Project (No.~2024QN11X088), and the Guangzhou Science and Technology
Planning Project (No.~2025A03J4491).

%% file: sections/appendix.tex
\section{Dataset Construction Details}
\label{app:datasets}

\subsection{Flickr30K}
We use the standard \textbf{Flickr30K}~\cite{Plummer2015Flickr30k} dataset, containing 31,783 images collected from Flickr, each associated with 5 human-annotated captions. We follow the standard split defined by~\citet{karpathy2015deep}, using 29,000 images for training, 1,014 for validation, and 1,000 for testing. All images are resized to $224 \times 224$ resolution during training and inference. Captions are tokenized using the standard CLIP tokenizer with a maximum sequence length of 77.

\subsection{COCO30K Sampling Protocol}
To ensure a fair comparison with Flickr30K-trained models, we construct \textbf{COCO30K}, a subset of the MS-COCO~\cite{Lin2014COCO} dataset aligned with Flickr30K in terms of training scale. The construction protocol is as follows:
\begin{itemize}[leftmargin=*]
    \item \textbf{Source:} We sample from the MS-COCO 2014 training split.
    \item \textbf{Sampling Unit:} Sampling is performed at the image level to preserve the original image--caption pairing (5 captions per image).
    \item \textbf{Scale Alignment:} We uniformly sample \textbf{29,000 images} to exactly match the number of training images in Flickr30K. This results in approximately 145,000 training captions.
    \item \textbf{Random Seed:} The sampling is performed with a fixed random seed (\texttt{seed=42}) to ensure reproducibility.
    \item \textbf{Validation/Test:} We evaluate models trained on COCO30K using the standard MS-COCO 5K test split. No images from the test split are included in the training subset.
\end{itemize}

\subsection{Dataset Summary Statistics}
Table~\ref{tab:dataset_stats} summarizes the split sizes and preprocessing settings of the two datasets.

\begin{table}[htbp]
\centering
\resizebox{\columnwidth}{!}{%
\begin{tabular}{lccccc}
\toprule
Dataset & \# Train & \# Val & \# Test & Caps/Img & Resolution \\
\midrule
Flickr30K & 29,000 & 1,014 & 1,000 & 5 & $224^2$ \\
COCO30K & 29,000 & 5,000 & 5,000 & 5 & $224^2$ \\
\bottomrule
\end{tabular}%
}
\caption{Summary of dataset statistics and preprocessing details.}
\label{tab:dataset_stats}
\end{table}

\section{Evaluation Protocol Details}
\label{app:eval}

\subsection{Retrieval Evaluation}
We report Recall@K (R@1, R@5, R@10) on the standard test sets.
\begin{itemize}[leftmargin=*]
    \item \textbf{Flickr30K:} Evaluated on the 1,000-image Karpathy test split.
    \item \textbf{MS-COCO:} Evaluated on the 5,000-image Karpathy test split.
\end{itemize}
Retrieval is performed by computing the cosine similarity between the query embedding and all candidate embeddings in the test set (Full Ranking).
\textbf{Candidate Pool Size:} For Flickr30K, the pool size is 1,000 images (or 5,000 texts). For MS-COCO, the pool size is 5,000 images (or 25,000 texts).

\section{Baseline Implementation Notes}
\label{app:baselines}

We compare our method against representative baselines covering hard-negative mining, language augmentation, and visual synthesis. All baselines share the same backbone architecture and are trained under the unified budget described in Appendix~\ref{app:hyperparams}. Crucially, for all baselines involving image generation or synthesis (ALiA, COCO-CF, TripletCLIP, and composed baselines), we utilize the same \textbf{Flux2-9B} model with identical generation parameters (8 inference steps, $512 \times 512$ resolution) to ensure a strictly fair comparison of the underlying methodologies.

\begin{itemize}[leftmargin=*]
    \item \textbf{Vanilla:} The backbone model is fine-tuned using the standard CLIP objective (InfoNCE loss) with random in-batch negatives.
    \item \textbf{NegCLIP:} We implement the NegCLIP strategy by constructing hard negative captions through rule-based compositional perturbation of the original caption. Following the official protocol, we generate up to 5 hard-negative variants per caption; captions without sufficient hard negatives are skipped, yielding 205K variants in total.
    \item \textbf{SugarCrepe-style:} We adopt the hard-negative generation strategy from SugarCrepe. We use a Large Language Model (LLM) to generate compositional hard negatives (e.g., swapping objects, changing attributes) for the training captions. We generate 1 hard-negative variant per caption.
    \item \textbf{LaCLIP:} We implement LaCLIP by generating rewritten captions using an LLM. During training, we randomly sample either the original caption or a rewritten version to encourage invariance to linguistic variations. We generate 1 rewritten variant per caption.
    \item \textbf{ALiA:} Following ALiA, we generate synthetic images using a text-to-image diffusion model (Flux2-9B, 8 steps, $512 \times 512$). The prompts are generated by an LLM to encompass counterfactual scenarios or attribute changes. These synthetic images serve as data augmentation. We generate 5 synthetic variants per image.
    \item \textbf{COCO-CF \& TripletCLIP:} We utilize the provided counterfactual data and training strategies from COCO-CF and TripletCLIP, incorporating their respective synthetic negative pairs into the contrastive loss. For each sample, we generate 1 caption variant and 1 corresponding synthetic image (using Flux2-9B with identical settings).
    \item \textbf{LaCLIP + SugarCrepe (Composed Baseline):} To ensure a fair comparison with our RePair framework, we combine LaCLIP and SugarCrepe. Specifically, we apply LaCLIP's text rewriting augmentation while simultaneously employing SugarCrepe's hard-negative generation capability. We carefully tune the ratio of rewritten texts to ensure the total number of training pairs seen by the model aligns with our computational budget.
    \item \textbf{COCO-CF + ALiA + LaCLIP (Composed Baseline):} We further evaluate a stronger naive composition that aggregates (i) COCO-CF's counterfactual negative pairs ($I^{-},T^{-}$), (ii) ALiA's diffusion-based image augmentation ($I^{+}$), and (iii) LaCLIP's rewrites ($T^{+}$). We construct the composed training set by merging the three sources and sampling to match the total sample count reported in the Samples~(k) column of Table~\ref{tab:merged_results}, without additional error-aware filtering or model-in-the-loop conditioning. All image synthesis is performed using the shared Flux2-9B setup.

\end{itemize}

\section{Full Training Hyperparameters}
\label{app:hyperparams}

All experiments are conducted using the PyTorch framework. We initialize the dual-encoder backbone with the pre-trained CLIP weights (ViT-B/32). To ensure optimal performance, we performed a grid search for key hyperparameters; the optimal configuration for each dataset was selected from the ranges listed below. The default training configuration and search spaces are detailed in Table~\ref{tab:hyperparams}.

\begin{table}[htbp]
\centering
\resizebox{\columnwidth}{!}{%
\begin{tabular}{l l}
\toprule
\textbf{Hyperparameter} & \textbf{Value / Search Range} \\
\midrule
\multicolumn{2}{l}{\textit{Optimization}} \\
Optimizer & AdamW \\
Learning Rate & \{1e-5, 5e-6, 2e-6, 1e-6\} \\
Weight Decay & 0.1 \\
LR Scheduler & Cosine Annealing \\
Warmup Ratio & 0.1 \\
\midrule
\multicolumn{2}{l}{\textit{Training Loop}} \\
Batch Size & \{64, 128, 256, 512, 1024\} \\
Training Epochs & \{3, \dots, 10\} \\
Precision & Mixed Precision (FP16) \\
\midrule
\multicolumn{2}{l}{\textit{Loss \& Regularization}} \\
$\lambda$ (Grid Loss Weight) & \{0.1, 0.2, 0.4, 0.5, 0.6, 0.8, 1.0\} \\
Seed & Average over 3 seeds \\
\midrule
\multicolumn{2}{l}{\textit{Image Generation (Flux2-9B)}} \\
Diffusion Model & Flux2-9B \\
Inference Steps & 8 \\
Output Resolution & $512 \times 512$ \\
\bottomrule
\end{tabular}%
}
\caption{Hyperparameter Settings and Grid Search Ranges.}
\label{tab:hyperparams}
\end{table}

\section{Algorithm and Notation Details}
\label{app:algorithms}

This section provides a comprehensive reference for the algorithmic details and mathematical notation used throughout the paper. We first present a summary of the notation in Table~\ref{tab:notation}. Subsequently, we provide the formal pseudocode for our failure-driven counterfactual synthesis pipeline (Algorithm~\ref{alg:synthesis}) and the $3 \times 3$ grid training procedure (Algorithm~\ref{alg:training}).

\begin{table}[!t]
\centering
\small
\begin{tabular}{l p{0.55\columnwidth}}
\toprule
Symbol & Meaning \\
\midrule
$\mathcal{D}=\{(I_i,T_i)\}_{i=1}^N$ & Paired image--text dataset \\
$I, T$ & An image / a text (generic) \\
$z_I(I), z_T(T)$ & $\ell_2$-normalized embeddings \\
$s(I,T)= z_I(I)^\top z_T(T)$ & Cosine similarity \\
$\ell(I,T)=\alpha\, s(I,T)$ & Training/retrieval logit \\

$\alpha$ (or $\tau = 1/\alpha$) & Logit scale (or temperature) \\

\midrule
$q$ & Query (either $T$ for T2I or $I$ for I2T) \\
$\mathcal{A}$ & Query set of a retrieval direction \\
$p$ & A candidate item scored against $q$ \\
$\mathcal{C}(q), \mathcal{P}(q)$ & Cross-modal candidate set / annotated positives of $q$ \\
$\operatorname{Top}^{n}$ & The $n$ candidates with the highest scores \\
$K_{\mathrm{mine}}$ & Size of the mined false-positive pool \\
$\mathcal{F}(q)$ & False-positive pool of $q$ (Eq.~\ref{eq:fp_pool_image}) \\
$d$ & False positives retained per query (Eq.~\ref{eq:hard_fp}) \\
$p^{\mathrm{gt}}$ & Ground-truth paired match of $q$ \\
$p^{\mathrm{fp}}$ & Retrieved false positive (FP) for $q$ \\
$V_{\mathrm{GT}}, V_{\mathrm{FP}}\in\{0,1\}$ & QC verifiers: query--GT consistency (Criterion~1) and query--FP validity (Criterion~2) \\
\midrule
$G(q,p^{\mathrm{fp}})$ & Instruction generator (LLM) \\
$\tau_{\mathrm{edit}}$ & Edit instruction (output of $G$) \\
$E(p^{\mathrm{fp}},\tau_{\mathrm{edit}};\xi)$ & Editor with randomness $\xi$ \\
$\{p^{\mathrm{edit}}_m\}_{m=1}^M$ & $M$ edited candidates \\
$p^{\mathrm{edit}}$ & Selected counterfactual edit \\
\midrule
$I^{\mathrm{a}}, T^{\mathrm{a}}$ & Anchor matched pair in $3\times3$ grid \\
$I^{\mathrm{hn}}, T^{\mathrm{hn}}$ & Hard-negative image/text \\
$I^{+}(T^{\mathrm{hn}}), T^{+}(I^{\mathrm{hn}})$ & Positive counterparts for hard negatives \\
$\mathbf{L}\in\mathbb{R}^{N_I\times N_T}$ & Full logit matrix over batch \\
$\mathbf{S}^{(b)}\in\mathbb{R}^{3\times 3}$ &  the $3 \times 3$ logit matrix of grid $b$ \\
$\mathcal{P}^{\mathrm{a}}$ & Set of positive anchor pairs over batch\\
$\mathcal{G}$ & Set of the $B$ grids built from a batch\\
$\mathcal{Q}$ & Set of positive pairs in $3 \times 3$ local confusion grid\\
$\mathcal{J}^{(b)}_{ij}$ & Per-pair loss term within grid $b$ (Eq.~\ref{eq:block_loss}) \\
$\mathcal{L}_{\mathrm{global}}$ & Symmetric CLIP/InfoNCE loss \\
$\mathcal{L}_{\mathrm{grid}}$ & local 3-way grid loss \\
$\lambda$ & Weight for $\mathcal{L}_{\mathrm{grid}}$ \\
\bottomrule
\end{tabular}
\caption{Notation used in failure-driven counterfactual synthesis
and $3 \times 3$ grid training.}
\label{tab:notation}
\end{table}

\begin{algorithm}[htbp]
\caption{Failure-Driven Counterfactual Synthesis}
\label{alg:synthesis}
\begin{algorithmic}[1]
\REQUIRE Paired dataset $\mathcal{D}=\{(I_i,T_i)\}_{i=1}^N$; retrieval pool size $K_{\mathrm{mine}}$; retention counts $d_{\mathrm{I2T}},d_{\mathrm{T2I}}$; verifiers $V_{\mathrm{GT}},V_{\mathrm{FP}}$; instruction
generator $G$; editor $E$; number of edit candidates $M$.
\ENSURE Curated records $\mathcal{R}$, each containing 
$(q, p^{\mathrm{gt}}, p^{\mathrm{fp}}, p^{\mathrm{edit}})$.

\STATE $\mathcal{R} \leftarrow \emptyset$
\FOR{\textbf{each} direction $\mathrm{dir} \in \{\mathrm{T2I}, \mathrm{I2T}\}$}
    \STATE $\mathcal{A} \leftarrow$ query set of $\mathrm{dir}$;\quad $d \leftarrow d_{\mathrm{dir}}$
    \COMMENT{distinct captions for T2I, distinct images for I2T}
    \FOR{\textbf{each} query $q \in \mathcal{A}$}
        \STATE $p^{\mathrm{gt}} \leftarrow$ an annotated positive of $q$
        \IF{$V_{\mathrm{GT}}(q, p^{\mathrm{gt}}) = 0$}
            \STATE \textbf{continue} \COMMENT{Discard noisy supervision}
        \ENDIF
        \STATE $\mathcal{F}(q) \leftarrow \operatorname*{Top}_{p \in \mathcal{C}(q) \setminus \mathcal{P}(q)}^{\,K_{\mathrm{mine}}} \, \ell(q, p)$
        \COMMENT{Eq.~\ref{eq:fp_pool_image}}
        \STATE $\{p^{\mathrm{fp}}_j\}_{j=1}^{d} \leftarrow \operatorname*{Top}_{p \in \mathcal{F}(q)}^{\,d} \, \ell(q, p)$
        \COMMENT{Eq.~\ref{eq:hard_fp}}
        \FOR{\textbf{each} $p^{\mathrm{fp}} \in \{p^{\mathrm{fp}}_j\}_{j=1}^{d}$}
            \IF{$V_{\mathrm{FP}}(q, p^{\mathrm{fp}}) = 1$}
                \STATE \textbf{continue} \COMMENT{Skip benign mismatch}
            \ENDIF
            \STATE $\tau_{\mathrm{edit}} \leftarrow G(q, p^{\mathrm{fp}})$
            \COMMENT{Generate edit instruction}
            \STATE Generate candidates $\{p^{\mathrm{edit}}_m\}_{m=1}^M$ where
            $p^{\mathrm{edit}}_m \leftarrow E(p^{\mathrm{fp}}, \tau_{\mathrm{edit}}; \xi_m)$
            \STATE $p^{\mathrm{edit}} \leftarrow$
            $\arg\max_{1 \le m \le M}\operatorname{cos}\!\bigl(z(p^{\mathrm{fp}}),\, z(p^{\mathrm{edit}}_m)\bigr)$
            \COMMENT{Minimal-change selection}
            \STATE $\mathcal{R} \leftarrow \mathcal{R} \cup \{(q, p^{\mathrm{gt}}, p^{\mathrm{fp}}, p^{\mathrm{edit}})\}$
        \ENDFOR
    \ENDFOR
\ENDFOR
\RETURN $\mathcal{R}$
\end{algorithmic}
\end{algorithm}

\paragraph{Detailed Grid and Loss Definitions.}
For grid $b$, with image and text lists $\mathcal{I}_b,\mathcal{T}_b$ as in Eq.~\ref{eq:block_structure}, the logit matrix $\mathbf{S}^{(b)}\in\mathbb{R}^{3\times 3}$ is defined element-wise as
\begin{equation}
\begin{split}
S^{(b)}_{ij}&=\ell(\mathcal{I}_b[i],\mathcal{T}_b[j])\\
&=\alpha\, z_I(\mathcal{I}_b[i])^\top z_T(\mathcal{T}_b[j]),
\end{split}
\label{eq:block_logit}
\end{equation}
for $i,j\in\{1,2,3\}$.
The global contrastive loss is the standard symmetric InfoNCE loss computed exclusively over the original anchor pairs. Let $\mathbf{L}\in\mathbb{R}^{N_I\times N_T}$ denote the full logit matrix over a batch, where $L_{ij}=\ell(I_i^{\mathrm{a}},T_j^{\mathrm{a}})$, and let $\mathcal{P}^{\mathrm{a}}$ be the set of matched anchor pairs:
\begin{equation}
\begin{split}
\mathcal{L}_{\mathrm{global}}
=\frac{1}{2|\mathcal{P}^{\mathrm{a}}|}\sum_{(i,j)\in\mathcal{P}^{\mathrm{a}}}\Bigg(
&-\log\frac{e^{L_{ij}}}{\sum_{k=1}^{N_T}e^{L_{ik}}} \\
&-\log\frac{e^{L_{ij}}}{\sum_{k=1}^{N_I}e^{L_{kj}}}
\Bigg).
\end{split}
\label{eq:main_loss}
\end{equation}

\begin{algorithm}[htbp]
\caption{$3 \times 3$ Local Confusion Grid Construction and Dual-Objective Training}
\label{alg:training}
\begin{algorithmic}[1]
\REQUIRE Dataset $\mathcal{D}$; curated records $\mathcal{R}$ (for each anchor, provide $I^{\mathrm{hn}},T^{\mathrm{hn}}$ and selected edits $I^{+}(T^{\mathrm{hn}}),T^{+}(I^{\mathrm{hn}})$); scale $\alpha$; local weight $\lambda\ge 0$.
\ENSURE Updated embeddings $(z_I,z_T)$.

\STATE Sample a minibatch of $B$ anchor pairs $\{(I_b^{\mathrm{a}},T_b^{\mathrm{a}})\}_{b=1}^{B}$ from $\mathcal{D}$
\STATE Initialize grid set $\mathcal{G}\leftarrow \emptyset$

\FOR{$b=1,\ldots,B$}
  \STATE Retrieve from $\mathcal{R}$: $I_b^{\mathrm{hn}},\,T_b^{\mathrm{hn}},\,I_b^{+}(T_b^{\mathrm{hn}}),\,T_b^{+}(I_b^{\mathrm{hn}})$
  \STATE Set $\mathcal{I}_b \leftarrow [\,I_b^{\mathrm{a}},\, I_b^{+}(T_b^{\mathrm{hn}}),\, I_b^{\mathrm{hn}}\,]$
  \STATE Set $\mathcal{T}_b \leftarrow [\,T_b^{\mathrm{a}},\, T_b^{\mathrm{hn}},\, T_b^{+}(I_b^{\mathrm{hn}})\,]$
  \STATE $\mathcal{G}\leftarrow \mathcal{G}\cup\{(\mathcal{I}_b,\mathcal{T}_b)\}$
\ENDFOR

\STATE Compute anchor embeddings $\mathbf{Z}_I^{\mathrm{a}}=\{z_I(I_b^{\mathrm{a}})\}_{b=1}^{B}$,
$\mathbf{Z}_T^{\mathrm{a}}=\{z_T(T_b^{\mathrm{a}})\}_{b=1}^{B}$
\STATE Form $\mathbf{L}^{\mathrm{a}}=\alpha\,\mathbf{Z}_I^{\mathrm{a}}(\mathbf{Z}_T^{\mathrm{a}})^\top\in\mathbb{R}^{B\times B}$
\STATE $\mathcal{P}^{\mathrm{a}}\leftarrow \{(b,b)\}_{b=1}^{B}$
\STATE $\mathcal{L}_{\mathrm{global}}\leftarrow \frac{1}{2|\mathcal{P}^{\mathrm{a}}|}\sum_{(i,j)\in\mathcal{P}^{\mathrm{a}}}\Big(
-\log\frac{e^{L^{\mathrm{a}}_{ij}}}{\sum_{k=1}^{B}e^{L^{\mathrm{a}}_{ik}}}
-\log\frac{e^{L^{\mathrm{a}}_{ij}}}{\sum_{k=1}^{B}e^{L^{\mathrm{a}}_{kj}}}\Big)$

\STATE $\mathcal{L}_{\mathrm{grid}}\leftarrow 0$;\quad $\mathcal{Q}\leftarrow \{(1,1),(2,2),(3,3)\}$
\FOR{each grid $(\mathcal{I}_b,\mathcal{T}_b)\in\mathcal{G}$}
  \STATE Compute grid embeddings $\mathbf{z}_I^{(b)}=\{z_I(\mathcal{I}_b[i])\}_{i=1}^{3}$,
  $\mathbf{z}_T^{(b)}=\{z_T(\mathcal{T}_b[j])\}_{j=1}^{3}$
  \STATE Compute $S^{(b)}_{ij}=\alpha\,\mathbf{z}_I^{(b)}[i]^\top \mathbf{z}_T^{(b)}[j]\in\mathbb{R}^{3\times 3}$
  \STATE $\mathcal{J}^{(b)}\leftarrow \frac{1}{2|\mathcal{Q}|}\sum_{(i,j)\in\mathcal{Q}}\Big(
  -\log\frac{e^{S^{(b)}_{ij}}}{\sum_{k=1}^{3}e^{S^{(b)}_{ik}}}
  -\log\frac{e^{S^{(b)}_{ij}}}{\sum_{k=1}^{3}e^{S^{(b)}_{kj}}}\Big)$
  \STATE $\mathcal{L}_{\mathrm{grid}}\leftarrow \mathcal{L}_{\mathrm{grid}}+\mathcal{J}^{(b)}$
\ENDFOR
\STATE $\mathcal{L}_{\mathrm{grid}}\leftarrow \mathcal{L}_{\mathrm{grid}}/B$
\STATE $\mathcal{L}\leftarrow \mathcal{L}_{\mathrm{global}}+\lambda\mathcal{L}_{\mathrm{grid}}$
\STATE Update $(z_I,z_T)$ by minimizing $\mathcal{L}$
\end{algorithmic}
\end{algorithm}

\section{Full Experimental Results}
\label{app:full_results}

This section provides the complete retrieval results that complement the condensed comparison in the main paper (Table~\ref{tab:merged_results}).
Whereas the main text reports only R@1 and R@5 for brevity, Tables~\ref{tab:full_flickr30k} and \ref{tab:full_coco30k} below list the full metric suite---R@1, R@5, R@10, and Mean Reciprocal Rank (MRR)---for both Image-to-Text (I2T) and Text-to-Image (T2I) retrieval on Flickr30K and COCO30K, respectively.
These expanded results show that RePair's advantage over the baselines is concentrated in R@1 and MRR, and holds on both datasets.

\begin{table*}[t]
\centering

\small
\setlength{\tabcolsep}{4pt}
\renewcommand{\arraystretch}{1.08}

\begin{tabular}{lccccc cccc cccc}
\toprule
    \multirow{2}{*}{Method} &
    \multirow{2}{*}{Samples (k)} &
    \multicolumn{4}{c}{Aug. Type} &
    \multicolumn{4}{c}{I2T} &
    \multicolumn{4}{c}{T2I} \\
    \cmidrule(lr){3-6} \cmidrule(lr){7-10} \cmidrule(lr){11-14}
    & & $I^{+}$ & $I^{-}$ & $T^{+}$ & $T^{-}$
    & R@1 & R@5 & R@10 & MRR
    & R@1 & R@5 & R@10 & MRR \\
    \midrule
    Vanilla
    & -- & & & & &
    87.87 & 98.22 & 99.30 & 0.9247 &
    74.60 & 92.87 & 96.21 & 0.8262 \\
    
    NegCLIP
    & 205 & & & & \cmark &
    88.07 & 98.13 & \best{99.51} & 0.9235 &
    74.38 & 92.78 & 96.19 & 0.8264 \\
    
    SugarCrepe
    & 145 & & & & \cmark &
    88.95 & \second{98.62} & \best{99.51} & 0.9273 &
    74.64 & 92.98 & 96.25 & 0.8284 \\
    
    LaCLIP
    & 145 & & & \cmark & &
    86.69 & 97.83 & 99.21 & 0.9157 &
    74.46 & 92.60 & 96.13 & 0.8257 \\
    
    SimCLR
    & 145 & \cmark & & & &
    88.26 & 98.32 & \second{99.41} & 0.9245 &
    74.30 & 92.78 & 96.11 & 0.8258 \\
    
    StableRep
    & 145 & \cmark & & & &
    87.67 & 98.22 & \second{99.41} & 0.9233 &
    74.64 & 92.72 & 96.02 & 0.8276 \\
    
    ALiA
    & 145 & \cmark & & & &
    87.67 & 98.22 & 99.21 & 0.9214 &
    74.36 & 92.88 & 96.29 & 0.8258 \\
    
    COCO-CF
    & 145 & & \cmark & & \cmark &
    87.97 & 97.44 & 99.21 & 0.9198 &
    74.64 & 92.98 & 96.45 & 0.8275 \\
    
    TripletCLIP
    & 145 & & \cmark & & \cmark &
    85.11 & 97.44 & 98.52 & 0.9035 &
    73.47 & 92.13 & 95.60 & 0.8169 \\
    \midrule
    \multicolumn{14}{l}{\textit{Combined Baselines}} \\
    La + SC
    & 290 & & & \cmark & \cmark &
    88.07 & 97.34 & 98.82 & 0.9212 &
    74.24 & 92.62 & 96.09 & 0.8276 \\

    CF + AL + La
    & 435 & \cmark & \cmark & \cmark & \cmark &
    86.88 & 97.63 & 98.72 & 0.8970 &
    74.14 & 92.84 & 96.15 & 0.8097 \\

    \midrule
    
    RePair-I2T
    & 47 & & & \cmark & \cmark &
    88.75 & 98.12 & 99.21 & \best{0.9350} &
    \second{76.07} & \best{93.73} & 96.42 & \best{0.8383} \\
    
    RePair-T2I
    & 60 & \cmark & \cmark &  &  &
    \second{89.44} & 98.32 & 99.21 & \second{0.9325} &
    75.93 & 93.47 & \second{96.52} & 0.8355 \\
    
    RePair (Full)
    & 107 & \cmark & \cmark & \cmark & \cmark &
    \best{90.13} & \best{98.71} & 99.30 & 0.9286 &
    \best{76.35} & \second{93.67} & \best{96.54} & \second{0.8370} \\

\bottomrule
\end{tabular}
\caption{Full retrieval performance on Image-to-Text (I2T) and Text-to-Image (T2I) on \textbf{Flickr30K}. The best and second-best results are marked in \textbf{bold} and \underline{underlined}, respectively. Augmentation modality and polarity are marked by $I^{+}$/$I^{-}$ (image hard positive/negative) and $T^{+}$/$T^{-}$ (text hard positive/negative). Combined baseline abbreviations: La\,=\,LaCLIP, SC\,=\,SugarCrepe, CF\,=\,COCO-CF, AL\,=\,ALiA.}
\label{tab:full_flickr30k}
\end{table*}

\begin{table*}[t]
\centering
\small
\setlength{\tabcolsep}{4pt}
\renewcommand{\arraystretch}{1.08}

\begin{tabular}{lccccc cccc cccc}
\toprule
    \multirow{2}{*}{Method} &
    \multirow{2}{*}{Samples (k)} &
    \multicolumn{4}{c}{Aug. Type} &
    \multicolumn{4}{c}{I2T} &
    \multicolumn{4}{c}{T2I} \\
    \cmidrule(lr){3-6} \cmidrule(lr){7-10} \cmidrule(lr){11-14}
    & & $I^{+}$ & $I^{-}$ & $T^{+}$ & $T^{-}$
    & R@1 & R@5 & R@10 & MRR
    & R@1 & R@5 & R@10 & MRR \\
    \midrule
    Vanilla
    & -- & & & & &
    60.54 & 83.86 & 90.94 & 0.7070 &
    44.15 & 71.97 & 81.74 & 0.5679 \\
    
    NegCLIP
    & 205 & & & & \cmark &
    61.06 & 84.00 & 91.24 & 0.7134 &
    44.37 & 72.29 & 81.95 & 0.5710 \\
    
    SugarCrepe
    & 145 & & & & \cmark &
    61.92 & 85.02 & 91.14 & 0.7167 &
    45.09 & 72.94 & 82.69 & 0.5778 \\
    
    LaCLIP
    & 145 & & & \cmark & &
    60.96 & 84.30 & 91.00 & \second{0.7267} &
    44.99 & 72.70 & 82.32 & 0.5778 \\
    
    SimCLR
    & 145 & \cmark & & & &
    61.34 & 84.14 & 90.86 & 0.7190 &
    45.13 & 72.77 & 82.24 & 0.5775 \\
    
    StableRep
    & 145 & \cmark & & & &
    61.76 & 84.98 & 91.02 & 0.7196 &
    44.61 & 72.95 & 82.47 & 0.5765 \\
    
    ALiA
    & 145 & \cmark & & & &
    61.62 & 84.40 & 90.58 & 0.7212 &
    44.93 & 72.87 & 82.38 & 0.5773 \\
    
    COCO-CF
    & 145 & & \cmark & & \cmark &
    60.32 & 84.92 & \second{91.60} & 0.7192 &
    44.55 & 72.61 & 82.51 & 0.5773 \\
    
    TripletCLIP
    & 145 & & \cmark & & \cmark &
    56.06 & 84.40 & 90.96 & 0.7183 &
    43.20 & 71.23 & 82.63 & 0.5780 \\
    \midrule
    \multicolumn{14}{l}{\textit{Combined Baselines}} \\
    La + SC
    & 290 & & & \cmark & \cmark &
    60.10 & 84.96 & 91.46 & 0.7225 &
    44.57 & 72.63 & 82.55 & 0.5816 \\

    CF + AL + La
    & 435 & \cmark & \cmark & \cmark & \cmark &
    59.74 & 83.98 & 91.04 & 0.7045 &
    44.56 & 72.59 & 81.49 & 0.5586 \\

    \midrule
    
    RePair-I2T
    & 47 & & & \cmark & \cmark &
    \second{62.06} & 84.84 & 91.10 & 0.7224 &
    45.69 & 73.31 & 82.61 & 0.5826 \\
    
    RePair-T2I
    & 60 & \cmark & \cmark &  &  &
    61.78 & \second{85.24} & 91.32 & 0.7217 &
    \second{45.87} & \second{73.53} & \second{82.93} & \second{0.5841} \\
    
    RePair (Full)
    & 107 & \cmark & \cmark & \cmark & \cmark &
    \best{62.64} & \best{85.72} & \best{91.64} & \best{0.7271} &
    \best{46.11} & \best{73.59} & \best{83.11} & \best{0.5863} \\

\bottomrule
\end{tabular}

\caption{Full retrieval performance on Image-to-Text (I2T) and Text-to-Image (T2I) on \textbf{COCO30K}. The best and second-best results are marked in \textbf{bold} and \underline{underlined}, respectively. Augmentation modality and polarity are marked by $I^{+}$/$I^{-}$ (image hard positive/negative) and $T^{+}$/$T^{-}$ (text hard positive/negative). Combined baseline abbreviations: La\,=\,LaCLIP, SC\,=\,SugarCrepe, CF\,=\,COCO-CF, AL\,=\,ALiA.}
\label{tab:full_coco30k}
\end{table*}

\section{\texorpdfstring{Qualitative Examples of $3 \times 3$ Grid Samples}{Qualitative Examples of 3x3 Grid Samples}}
\label{app:qualitative_blocks}

To concretely illustrate the $3 \times 3$ grid structure defined in Eq.~\ref{eq:block_structure}, we present two real-world examples from our experiments on Flickr30K and COCO30K. Each grid is constructed around a failure anchor $(I^{\mathrm{a}}, T^{\mathrm{a}})$---a ground-truth pair where the model initially retrieved a high-ranking false positive. The grid incorporates these hard negatives along with their synthesized or paired positive counterparts to form a dense, discriminative neighborhood.

\begin{figure*}[t]
\centering
\includegraphics[width=\textwidth]{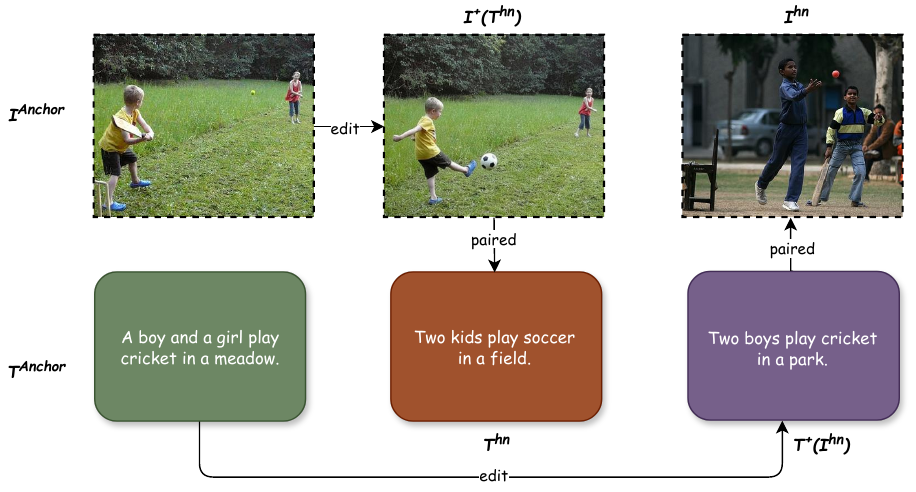}
\caption{A representative $3 \times 3$ grid from Flickr30K centered on an outdoor sports scenario. The anchor pair $(I^{\mathrm{a}}, T^{\mathrm{a}})$ depicts ``a boy and a girl play cricket,'' but the model retrieves a hard-negative text $T^{\mathrm{hn}}$ describing ``two kids play soccer.'' RePair synthesizes a corrected counterpart $T^{+}(I^{\mathrm{hn}})$ (``Two boys play cricket'') and retrieves the matched image $I^{+}(T^{\mathrm{hn}})$, creating a grid that forces the model to distinguish between fine-grained activities (cricket vs.\ soccer) and subjects (boy/girl vs.\ two boys).}
\label{fig:33_case_flickr}
\end{figure*}

In the Flickr30K example (Figure~\ref{fig:33_case_flickr}), the confusion boundary lies in the specific activity and subject composition. The hard negative $T^{\mathrm{hn}}$ shares the high-level semantic scaffold of ``children playing outdoors'' but hallucinates the sport (\textit{soccer} instead of \textit{cricket}) and generalizes the subjects (\textit{two kids} instead of \textit{boy and girl}). By including the edited counterpart $T^{+}(I^{\mathrm{hn}})$, which keeps the ``cricket'' context but aligns the subject description to the hard-negative image, the grid enforces a multi-axis discrimination task: the model must simultaneously verify the activity type and the subject demographics to maximize the diagonal scores.

\noindent The two edit instructions---one per direction---that seed this grid are:

\begin{tcolorbox}[
    colback=teal!5!white, colframe=teal!60!black,
    title={\small Flickr30K I2T Edit Instruction},
    fonttitle=\bfseries\small, arc=1.5mm, boxrule=0.6pt,
    left=4pt, right=4pt, top=2pt, bottom=2pt
]
\small
\texttt{"edit\_instruction"}: \textit{``[Action Error, Object Error]: The image shows a boy and a girl playing cricket in a grassy meadow, not two kids playing soccer in a field. Therefore, replace `Two kids' with `A boy and a girl', replace `soccer' with `cricket', and replace `in a field' with `in a meadow'.''}
\end{tcolorbox}

\begin{tcolorbox}[
    colback=orange!5!white, colframe=orange!60!black,
    title={\small Flickr30K T2I Edit Instruction},
    fonttitle=\bfseries\small, arc=1.5mm, boxrule=0.6pt,
    left=4pt, right=4pt, top=2pt, bottom=2pt
]
\small
\texttt{"edit\_instruction"}: \textit{``Change the two boys to a boy and a girl. Replace the park background with a natural green meadow without paths or benches, keeping the cricket activity and overall outdoor composition.''}
\end{tcolorbox}

\begin{figure*}[t]
\centering
\includegraphics[width=\textwidth]{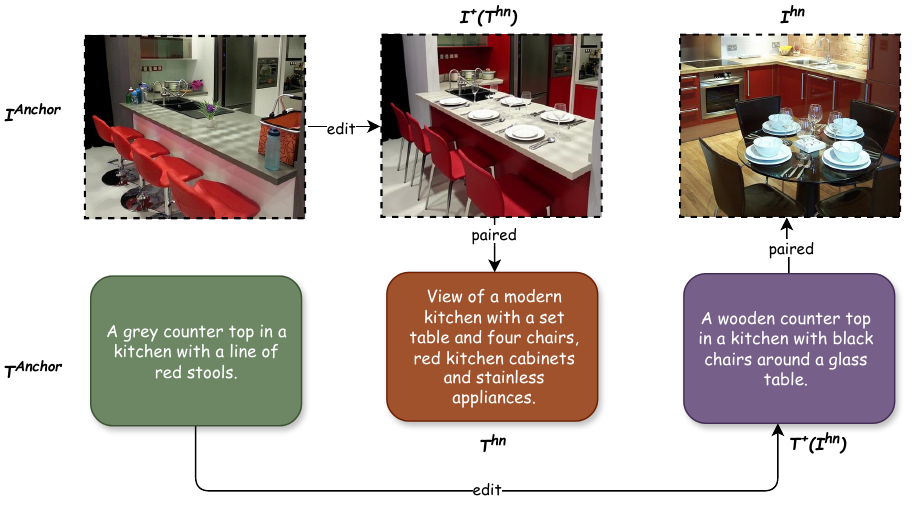}
\caption{A $3 \times 3$ grid from COCO30K capturing a kitchen scene. The anchor $(I^{\mathrm{a}}, T^{\mathrm{a}})$ features a ``grey counter top'' with ``red stools,'' while the hard negative $T^{\mathrm{hn}}$ describes a ``modern kitchen'' with ``red cabinets'' and ``stainless appliances.'' The synthesized positive $T^{+}(I^{\mathrm{hn}})$ corrects the attributes to match the hard-negative image $I^{\mathrm{hn}}$ (``wooden counter top,'' ``black chairs''), ensuring that the model learns to disentangle compositional attributes like color and material within the shared kitchen domain.}
\label{fig:33_case_coco}
\end{figure*}

Similarly, the COCO30K example (Figure~\ref{fig:33_case_coco}) illustrates a failure driven by attribute binding. All images in the grid depict kitchens, resulting in high baseline similarity. The confusion arises because the model mixes up specific attributes: \textit{red} applies to \textit{stools} in the anchor but to \textit{cabinets} in the hard negative. RePair's counterfactual synthesis produces a hard positive $T^{+}(I^{\mathrm{hn}})$ that precisely describes the \textit{wooden counter} and \textit{black chairs} of the hard-negative image $I^{\mathrm{hn}}$. This forces the embedding space to resolve the binding of color (\textit{red} vs.\ \textit{black}) and object (\textit{stools} vs.\ \textit{chairs}) rather than relying on the generic ``kitchen'' scene detection.

\noindent The two edit instructions---one per direction---that seed this grid are:

\begin{tcolorbox}[
    colback=teal!5!white, colframe=teal!60!black,
    title={\small COCO30K I2T Edit Instruction},
    fonttitle=\bfseries\small, arc=1.5mm, boxrule=0.6pt,
    left=4pt, right=4pt, top=2pt, bottom=2pt
]
\small
\texttt{"edit\_instruction"}: \textit{``[Object Error, Attribute Error]: The image shows a grey counter top with a row of red stools, not a table with four chairs. The red color applies to bar stools along the counter, not to kitchen cabinets. Therefore, replace `a set table and four chairs' with `a counter top with a line of red stools', and remove `red kitchen cabinets and stainless appliances'.''}
\end{tcolorbox}

\begin{tcolorbox}[
    colback=orange!5!white, colframe=orange!60!black,
    title={\small COCO30K T2I Edit Instruction},
    fonttitle=\bfseries\small, arc=1.5mm, boxrule=0.6pt,
    left=4pt, right=4pt, top=2pt, bottom=2pt
]
\small
\texttt{"edit\_instruction"}: \textit{``Replace the wooden counter top with a grey counter top. Remove the black chairs and glass table, and add a row of red bar stools along the counter. Maintain the kitchen environment, lighting, and overall layout.''}
\end{tcolorbox}

Both cases demonstrate the core advantage of the RePair framework: hard negatives are not random samples but genuine retrieval failures, and the synthesized positives are minimal edits designed to be maximally confusable yet semantically distinct. This structured $3 \times 3$ grid supervision concentrates the gradient on the most challenging boundaries of the manifold.

\section{Statistical Analysis of RePair Synthetic Samples}
\label{app:synthesis_stats}

This section provides a statistical characterization of the synthetic data generated by the RePair pipeline. We analyze the distribution of synthesized hard positives (HP) and mined hard negatives (HN) across image and caption anchors to understand the coverage and structural properties of the failure-driven augmentation.

\begin{figure*}[t]
\centering
\includegraphics[width=\textwidth]{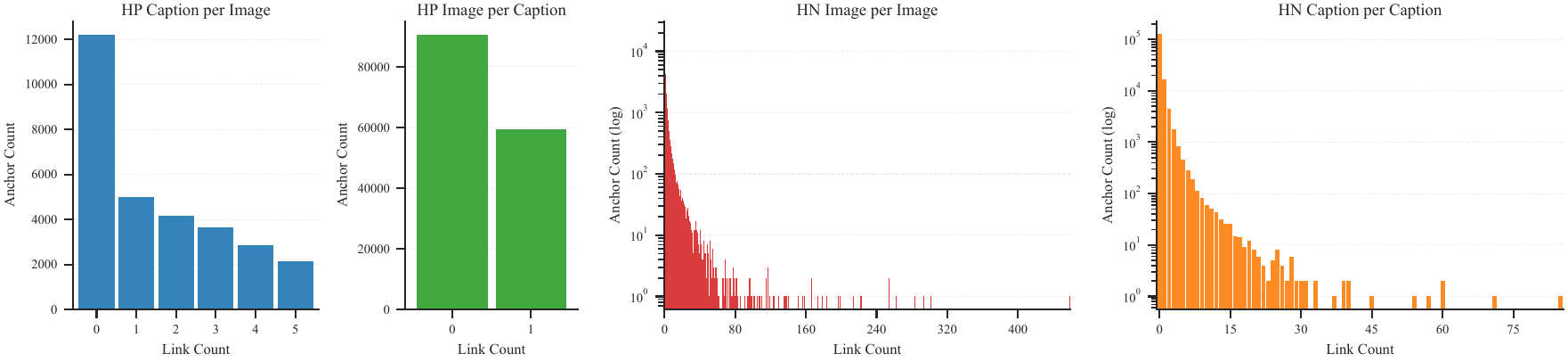}
\caption{Distribution of synthesized and mined samples for COCO30K. The four panels show: (Top-Left) Number of hard-positive captions generated per image anchor; (Top-Right) Number of hard-positive images generated per caption anchor; (Bottom-Left) Distribution of hard-negative images linked to each image anchor; (Bottom-Right) Distribution of hard-negative captions linked to each caption anchor. Note the long-tail distribution in hard negatives (log scale) versus the controlled sparsity of hard positives.}
\label{fig:coco30k_distributions}
\end{figure*}

\begin{figure*}[t]
\centering
\includegraphics[width=\textwidth]{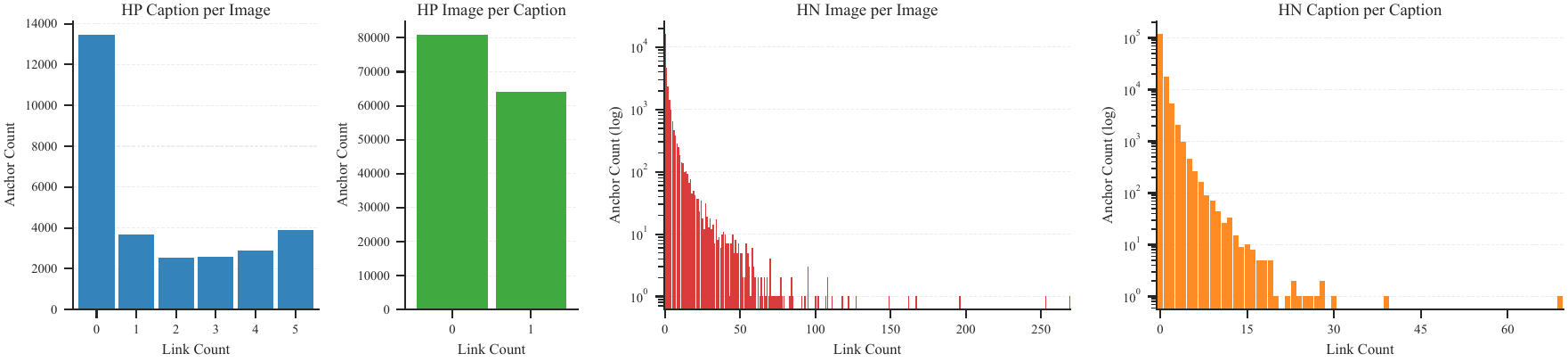}
\caption{Distribution of synthesized and mined samples for Flickr30K. The layout follows Figure~\ref{fig:coco30k_distributions}. While the overall patterns are similar to COCO30K, the tail of the hard-negative distribution is shorter (max $\sim$250 vs.\ $\sim$400).}
\label{fig:flickr30k_distributions}
\end{figure*}

\paragraph{Hard-Positive Distribution.}
The top row of Figures~\ref{fig:coco30k_distributions} and \ref{fig:flickr30k_distributions} reveals a distinct asymmetry between modalities. For I2T synthesis (Top-Left), most image anchors receive between 1 and 5 hard-positive captions, consistent with the retention count $d_{\text{I2T}}{=}5$ (Eq.~\ref{eq:hard_fp}). In contrast, for T2I synthesis (Top-Right), the distribution is strictly binary or sparse, with most caption anchors receiving exactly one hard-positive image, reflecting $d_{\text{T2I}}{=}1$ and the higher computational cost of image generation.

\paragraph{Hard-Negative Distribution.}
The bottom row displays the connectivity of mined hard negatives on a logarithmic scale. Both datasets exhibit a heavy-tailed distribution: while the median anchor is associated with a moderate number of hard negatives (typically 10--50), a small fraction of ``hub'' anchors attract a disproportionately large number of negatives (up to $\sim$250 for Flickr30K and $\sim$400 for COCO30K). These hubs correspond to semantically dense regions of the embedding space---such as generic ``kitchen'' scenes or ``people playing sports''---where many distinct instances share high visual-semantic overlap, making them frequent false positives for numerous queries.

\paragraph{Cross-Dataset Comparison.}
Comparing the two datasets, COCO30K exhibits wider tails in its hard-negative distributions (Bottom-Left/Right) than Flickr30K. This is attributable to the broader diversity of scenes in MS-COCO. However, the hard-positive distributions (Top-Left/Right) remain remarkably consistent across both datasets. This stability suggests that the synthesis pipeline's behavior is governed more by the model's error rate and the synthesis protocol than by the underlying dataset size, ensuring scalable performance.

\paragraph{Implications.}
The combination of sparse, high-precision hard positives and dense, long-tail hard negatives validates the ``precision over volume'' philosophy of RePair. Rather than uniformly augmenting every training sample, the framework concentrates its generative capacity on the specific anchors that sit at the intersection of multiple confusion boundaries (the hubs), thereby efficiently densifying the supervision where the model is most uncertain. 

\section{Influence of Synthetic Sample Size}
\label{app:sample_size}

\begin{figure*}[t]
\centering
\includegraphics[width=0.85\textwidth]{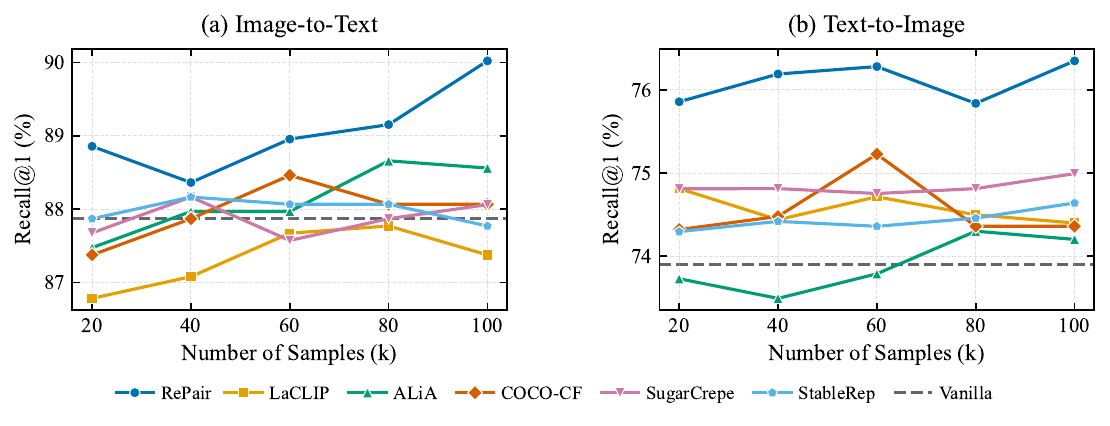}
\vspace{-10pt}
\caption{Effect of synthetic sample budget on Recall@1 for Flickr30K. (a)~Image-to-Text and (b)~Text-to-Image retrieval across 20k--100k synthesized samples. RePair consistently outperforms all baselines at every budget level, while error-agnostic methods saturate or regress as sample count grows.}
\label{fig:num_samples}
\end{figure*}

Figure~\ref{fig:num_samples} reports Recall@1 as the synthesis budget varies from 20k to 100k on Flickr30K. RePair traces the upper envelope of performance in both I2T and T2I across the entire range: it already delivers clear gains at low budgets (20--40k) and continues to improve steadily as more samples are added, exhibiting a predictable scaling curve. This confirms that error-driven counterfactual synthesis yields a higher density of useful supervision per sample (see Appendix~\ref{app:synthesis_stats} for a detailed statistical analysis of the synthesized samples).

In contrast, non-failure-driven baselines (LaCLIP, ALiA, COCO-CF, SugarCrepe, StableRep) exhibit early saturation or occasional regression at higher budgets, indicating that error-agnostic synthetic data can introduce redundant or misaligned training signals that offset marginal gains. The divergence between RePair and these baselines widens with budget size, reinforcing that \textit{targeting the model's actual error space} matters more than simply scaling synthetic volume.

\section{Synthesis Pipeline Efficiency Analysis}
\label{app:efficiency}

The RePair synthesis pipeline involves three LLM-based stages: (1) Quality Control, (2) Edit Instruction Generation, and (3) Editing (caption editing for I2T; image editing via Flux2-9B for T2I). Only stages involving LLM calls (Gemini 2.5 Flash) incur token costs; T2I image editing uses Flux2-9B and does not consume LLM tokens. The reported token usage is based on processing \textbf{1,000 mined failure cases per direction (I2T and T2I) for each dataset} (Flickr30K and COCO30K), a fixed measurement sample used for all cost comparisons. Table~\ref{tab:token_usage_appendix_i2t_t2i} reports token usage in thousands (K) broken down by dataset, direction, and stage.

\begin{table*}[t]
    \centering
    \renewcommand{\arraystretch}{1.05}
    \resizebox{\textwidth}{!}{%
    \begin{tabular}{ll
      S[table-format=4.1] S[table-format=4.1] S[table-format=4.1]
      S[table-format=4.1] S[table-format=4.1] S[table-format=4.1]
      S[table-format=4.1] S[table-format=4.1] S[table-format=4.1]
      S[table-format=4.1] S[table-format=4.1] S[table-format=4.1]}
    \toprule
    & & \multicolumn{3}{c}{Quality Control} & \multicolumn{3}{c}{Instruction Generation} & \multicolumn{3}{c}{Editing} & \multicolumn{3}{c}{Total} \\
    \cmidrule(lr){3-5}\cmidrule(lr){6-8}\cmidrule(lr){9-11}\cmidrule(lr){12-14}
    Dataset & Dir
    & {In} & {Out} & {Tot}
    & {In} & {Out} & {Tot}
    & {In} & {Out} & {Tot}
    & {In} & {Out} & {Tot} \\
    \midrule

\multirow{3}{*}{Flickr30K}
    & I2T & \tok{707.3} & \tok{80.3} & \tok{787.6} & \tok{981.4} & \tok{61.7} & \tok{1043.1} & \tok{597.9} & \tok{45.9} & \tok{643.8} & \tok{2286.6} & \tok{187.9} & \tok{2474.5} \\
    & T2I & \tok{1064.6} & \tok{101.0} & \tok{1165.6} & \tok{279.0} & \tok{25.8} & \tok{304.8} & \multicolumn{1}{c}{-} & \multicolumn{1}{c}{-} & \multicolumn{1}{c}{-} & \tok{1343.6} & \tok{126.8} & \tok{1470.4} \\
    & \textbf{Combined}
          & \tok{1771.9} & \tok{181.3} & \tok{1953.2} & \tok{1260.3} & \tok{87.5} & \tok{1347.8} & \tok{597.9} & \tok{45.9} & \tok{643.8} & \tok{3630.2} & \tok{314.7} & \tok{3944.9} \\
    \midrule
\multirow{3}{*}{COCO30K}
    & I2T & \tok{701.9} & \tok{69.5} & \tok{771.4} & \tok{706.7} & \tok{31.1} & \tok{737.8} & \tok{382.2} & \tok{25.2} & \tok{407.4} & \tok{1790.8} & \tok{125.8} & \tok{1916.6} \\
    & T2I & \tok{1063.3} & \tok{95.3} & \tok{1158.6} & \tok{214.9} & \tok{20.6} & \tok{235.5} & \multicolumn{1}{c}{-} & \multicolumn{1}{c}{-} & \multicolumn{1}{c}{-} & \tok{1278.2} & \tok{115.9} & \tok{1394.1} \\
    & \textbf{Combined}
          & \tok{1765.2} & \tok{164.8} & \tok{1930.0} & \tok{921.5} & \tok{51.8} & \tok{973.3} & \tok{382.2} & \tok{25.2} & \tok{407.4} & \tok{3069.0} & \tok{241.7} & \tok{3310.7} \\
    \bottomrule
    \end{tabular}%
    }
    \caption{Token usage (K tokens) by dataset, direction, and stage. ``Total'' sums over Stage 1--3; ``Combined'' sums I2T and T2I.}
    \label{tab:token_usage_appendix_i2t_t2i}
    \end{table*}

\paragraph{Per-Stage Analysis.}
The data reveals that \textbf{Quality Control} is the most token-intensive stage (Flickr30K: $\sim$1,953K; COCO30K: $\sim$1,930K combined), as it requires the LLM to assess both GT-query consistency and FP-query validity for every mined failure triplet. \textbf{Instruction Generation} is the second largest consumer (Flickr30K: $\sim$1,348K; COCO30K: $\sim$973K), reflecting the detailed error diagnosis and structured edit command generation. \textbf{Editing} (caption editing only; I2T direction) is the lightest stage (Flickr30K: $\sim$644K; COCO30K: $\sim$407K), as it only rewrites the false-positive caption following the already-generated instruction. T2I editing is handled by Flux2-9B and incurs zero LLM tokens (shown as ``-'' in the table).

\paragraph{Directional and Dataset Comparison.}
Comparing directions, I2T consumes more total LLM tokens than T2I (Flickr30K: 2,475K vs 1,470K) because I2T requires all three LLM stages, whereas T2I skips the text-based editing stage. T2I's image editing cost is instead reflected in GPU compute for Flux2-9B inference (8 steps, $512 \times 512$). Across datasets, Flickr30K and COCO30K exhibit very similar QC token usage ($\sim$1,953K vs $\sim$1,930K), as QC is run on the same fixed budget of 1,000 cases per direction for both datasets, with per-case prompts of comparable length. However, Flickr30K shows higher Instruction Generation and Editing costs than COCO30K, likely due to a higher yield of verified failures passing QC on Flickr30K.

\paragraph{Conclusion.}
These results underscore that RePair's failure-driven synthesis is not only data-efficient (107K samples vs.\ 145K--435K for baselines, a 26\%--75\% reduction) but also computationally inexpensive: using Gemini 2.5 Flash pricing, the total LLM cost is $\sim$\$1.88 for Flickr30K and $\sim$\$1.53 for COCO30K per 1K failure cases per direction. Since cost scales linearly with the number of processed failures, a full single round (47K I2T and 60K T2I instances) costs $\sim$\$98 and $\sim$\$80 respectively, keeping the approach practical for iterative refinement and multi-round synthesis at various scales.

\subsection{Unified Generation Overhead Comparison}
\label{app:cost_comparison}

To enable a fair, apple-to-apple comparison of generation overhead, we measure the total number of LLM API calls and image-generation invocations for each method under our unified experimental setting (same ViT-B/32 backbone, same Flux2-9B image generator). For baselines with native quality-control mechanisms (SugarCrepe, ALiA, COCO-CF), we apply a standardized 3-candidates-to-1-selected alignment matching RePair's $M{=}3$ post-edit selection protocol. Table~\ref{tab:cost_comparison} reports the results.

\begin{table*}[t]
\centering
\small
\setlength{\tabcolsep}{5pt}
\begin{tabular}{lcccccc}
\toprule
Method & QC? & Samples (K) & LLM (K) & Image (K) & LLM ratio & Image ratio \\
\midrule
NegCLIP & No & 205 & 0 & 0 & 0$\times$ & 0$\times$ \\
LaCLIP & No & 145 & 145 & 0 & 0.41$\times$ & 0$\times$ \\
TripletCLIP & No & 145 & 145 & 145 & 0.41$\times$ & 0.81$\times$ \\
SugarCrepe & Yes & 145 & 435--1,305 & 0 & 1.23--3.68$\times$ & 0$\times$ \\
ALiA & Yes & 145 & 145 & 435 & 0.41$\times$ & 2.42$\times$ \\
COCO-CF & Yes & 145 & 435 & 435 & 1.23$\times$ & 2.42$\times$ \\
CF+ALiA+LaCLIP & Partial & 435 & 725 & 870 & 2.04$\times$ & 4.83$\times$ \\
\midrule
RePair-I2T & Yes & 47 & 235 & 0 & 0.66$\times$ & 0$\times$ \\
RePair-T2I & Yes & 60 & 120 & 180 & 0.34$\times$ & 1.00$\times$ \\
\textbf{RePair (Full)} & Yes & \textbf{107} & \textbf{355} & \textbf{180} & \textbf{1.00$\times$} & \textbf{1.00$\times$} \\
\bottomrule
\end{tabular}
\caption{Unified generation overhead comparison. ``LLM (K)'' and ``Image (K)'' denote total LLM API calls and image-generation invocations in thousands. Baselines with native QC are aligned at 3-candidates-to-1-selected to match RePair's $M{=}3$ protocol. Ratios are relative to RePair (Full).}
\label{tab:cost_comparison}
\end{table*}

RePair's per-sample call count is detailed in Table~\ref{tab:repair_call_breakdown}: I2T requires 5 LLM calls per sample (1~QC + 1~instruction generation + 3~text edits for $M{=}3$ candidates), while T2I requires 2~LLM calls + 3~Flux image edits.

\begin{table}[htbp]
\centering
\small
\resizebox{\columnwidth}{!}{%
\begin{tabular}{lcccc}
\toprule
Variant & Samples & LLM/sample & Image/sample & Total LLM / Image \\
\midrule
RePair-I2T & 47K & 5 & 0 & 235K / 0 \\
RePair-T2I & 60K & 2 & 3 & 120K / 180K \\
\textbf{Full} & \textbf{107K} & --- & --- & \textbf{355K / 180K} \\
\bottomrule
\end{tabular}%
}
\caption{RePair per-sample call breakdown.}
\label{tab:repair_call_breakdown}
\end{table}

Four conclusions emerge from Table~\ref{tab:cost_comparison}. First, \textbf{RePair's LLM overhead is moderate}: 355K calls is lower than SugarCrepe (435K--1.3M), COCO-CF (435K), and the composite baseline CF+ALiA+LaCLIP (725K), and is higher only than methods that make at most one LLM call per sample (NegCLIP, LaCLIP, TripletCLIP, ALiA). Second, \textbf{RePair has the lowest image-generation cost among quality-controlled generation-based methods}: 180K invocations vs.\ 435K for ALiA and COCO-CF, and 870K for the composite baseline. Third, \textbf{higher total overhead does not translate to better performance}: CF+ALiA+LaCLIP consumes 2$\times$ the LLM calls and 4.8$\times$ the image calls of RePair, yet degrades below Vanilla (Table~\ref{tab:merged_results}: 86.88 vs.\ 87.87 Flickr30K I2T). This confirms that RePair's efficiency stems from precision-targeted synthesis, not volume. Fourth, \textbf{all pipeline complexity is offline}: training and inference are identical to vanilla CLIP fine-tuning, with zero deployment-time overhead.

\section{Validation of Synthesized Supervision}
\label{app:qc_eval}

This section provides a direct evaluation of the Quality Control (QC) module beyond the ablation evidence in Table~\ref{tab:ablation_r1}. We analyze (1)~the filtering statistics of each QC criterion across all experimental settings, (2)~agreement between QC decisions and human annotators, and (3)~verification by human annotators of the edited pairs that are retained for training.

\subsection{QC Filtering Statistics}

Our QC module applies two complementary criteria to each mined failure triplet $(q, p^{\mathrm{gt}}, p^{\mathrm{fp}})$ before synthesis:
\begin{itemize}[leftmargin=*]
    \item \textbf{Criterion~1 (Query--GT Consistency):} Verifies that the query $q$ and its paired ground truth $p^{\mathrm{gt}}$ are semantically consistent. Inconsistent pairs (e.g., a Flickr caption that poorly describes its image) are discarded to avoid noisy supervision.
    \item \textbf{Criterion~2 (Query--FP Validity):} Checks whether the false positive $p^{\mathrm{fp}}$ is already a valid alternative match for $q$ (a ``benign'' FP). If so, treating it as a hard negative would inject incorrect contrastive signal, so the triplet is skipped.
\end{itemize}

Table~\ref{tab:qc_filtering} reports the relative contribution of each criterion to the total filtered samples across all four settings.

\begin{table}[htbp]
\centering
\small
\setlength{\tabcolsep}{4pt}
\begin{tabular}{llcc}
\toprule
Dataset & Direction & Criterion~1 (\%) & Criterion~2 (\%) \\
\midrule
Flickr30K & T2I & 32.39 & \textbf{67.61} \\
Flickr30K & I2T & \textbf{57.61} & 42.39 \\
COCO30K & T2I & 25.85 & \textbf{74.15} \\
COCO30K & I2T & 27.93 & \textbf{72.07} \\
\bottomrule
\end{tabular}
\caption{Distribution of QC-filtered samples by criterion. Each row sums to 100\% of total filtered samples for that setting.}
\label{tab:qc_filtering}
\end{table}

Three observations emerge. First, \textbf{both criteria contribute non-trivially in every setting} (range: 25.85\%--74.15\%), confirming they capture different noise types. This is consistent with the ablation: removing the full QC ($-$0.66/$-$1.24) causes larger degradation than removing only post-edit selection ($-$0.52/$-$0.78), validating the complementary necessity of both criteria. Second, \textbf{benign FP filtering (Criterion~2) dominates T2I} (67.61\%--74.15\%). Retrieved FP images are often visually close to the query text---they appear ``roughly correct'' but lack fine-grained alignment. Without filtering, these benign FPs would dilute the contrastive training signal as false hard negatives. Third, \textbf{filtering behavior is consistent with dataset characteristics}: COCO30K shows systematically higher Criterion~2 rates than Flickr30K, reflecting its denser semantic space (80 object categories). Flickr30K I2T is the only setting where Criterion~1 dominates (57.61\%). These dataset-consistent patterns indicate that QC captures genuine quality signals rather than performing arbitrary filtering.

\subsection{Agreement between QC Decisions and Human Annotators}

To verify QC reliability against human annotators, we conduct an evaluation on COCO30K. We sample 200 QC-accepted and 200 QC-rejected triplets (total $N{=}400$), and ask human annotators to judge whether each QC decision is correct.

\begin{table}[htbp]
\centering
\small
\begin{tabular}{lc}
\toprule
Metric & Value \\
\midrule
Precision & 90.2\% \\
Recall & 83.9\% \\
Accuracy & 82.0\% \\
Cohen's $\kappa$ & 0.58 \\
\bottomrule
\end{tabular}
\caption{Agreement between QC decisions and human annotators ($N{=}400$, COCO30K).}
\label{tab:qc_human}
\end{table}

The high precision (90.2\%) indicates that the vast majority of QC-accepted samples are genuinely valid failures, ensuring synthesis quality. The slightly lower recall (83.9\%) reflects a conservative filtering strategy: QC occasionally rejects valid failures, but this is preferable to admitting noisy samples. Cohen's $\kappa{=}0.58$ indicates moderate agreement~\citep{landis1977measurement}, a strong result given the inherent subjectivity in judging whether a near-miss retrieval result is ``sufficiently similar'' to the query. Importantly, even without QC, RePair still leads all baselines on COCO30K I2T (61.98 vs.\ SugarCrepe 61.92), confirming that QC raises the performance ceiling but is not a fundamental prerequisite.

\subsection{Post-Edit Verification by Human Annotators}
\label{app:post_edit_audit}

The study above validates the \emph{input} to synthesis---whether a mined triplet is a genuine retrieval failure---but not its \emph{output}. We therefore ask human annotators to re-verify the edited pairs retained for training.

\paragraph{Protocol.} We sample 800 retained post-edit records, stratified as 200 per dataset $\times$ direction split (COCO30K I2T, COCO30K T2I, Flickr30K I2T, Flickr30K T2I). Annotators review records on side-by-side sheets and judge each on five binary criteria: (1)~the original false positive is genuinely wrong for the query; (2)~the edited sample matches the query; (3)~the shared scaffold is preserved; (4)~the edit is minimal, i.e.\ confined to the failure-causing residual; and (5)~the original and edited items form a valid hard contrast.

\begin{table}[htbp]
\centering
\small
\begin{tabular}{lccc}
\toprule
Criterion & Overall & I2T & T2I \\
\midrule
Original FP genuinely wrong & 99.5 & 99.0 & 100.0 \\
Edited sample matches query & 95.8 & 96.5 & 95.0 \\
Scaffold preserved & 97.0 & 99.3 & 94.8 \\
Edit is minimal & 93.3 & 96.8 & 89.8 \\
Valid hard contrast & \textbf{92.5} & \textbf{95.5} & \textbf{89.5} \\
\bottomrule
\end{tabular}
\caption{Verification of retained post-edit pairs by human annotators (\%; $N{=}800$, 200 per dataset $\times$ direction split).}
\label{tab:post_edit_audit}
\end{table}

\begin{table}[htbp]
\centering
\small
\begin{tabular}{lcc}
\toprule
 & COCO30K & Flickr30K \\
\midrule
I2T (text edit) & 98.5 & 92.5 \\
T2I (image edit) & 92.0 & 87.0 \\
\bottomrule
\end{tabular}
\caption{Valid hard-contrast rate (\%) by dataset and direction.}
\label{tab:post_edit_2x2}
\end{table}

Overall, \textbf{92.5\% of retained pairs form valid hard contrasts}. Two patterns organize the remainder. First, quality is direction-dependent: the I2T--T2I gap is 6.0\,pp on the final criterion and concentrates in scaffold preservation ($-4.5$\,pp) and minimality ($-7.0$\,pp), the two criteria most sensitive to how much of an image a diffusion editor rewrites. Second, a dataset effect of comparable size acts independently (Table~\ref{tab:post_edit_2x2}): COCO30K exceeds Flickr30K by 6.0\,pp for I2T and 5.0\,pp for T2I, consistent with Flickr30K's greater scene and caption diversity. The best and worst splits differ by 11.5\,pp, with the two factors each accounting for roughly half of the spread. Appendix~\ref{app:edit_failure_modes} classifies the remaining 60 records.

Two further points connect this study to the rest of the pipeline. The 99.5\% rate on the first criterion exceeds the 90.2\% QC precision in Table~\ref{tab:qc_human} because the two are measured on different populations: Table~\ref{tab:qc_human} covers all QC-accepted triplets, whereas this study covers the narrower set for which an edit was produced and retained by post-edit selection. And the 93.3\% minimality rate corroborates the embedding-proximity criterion used for that selection---the cosine-similarity proxy in Table~\ref{tab:edit_quality_by_type} and judgments from human annotators independently indicate that retained edits stay confined to the residual.

\section{Failure Characterization}
\label{app:failure_char}

RePair does not treat all retrieval failures as uniform. In a dedicated analysis pass over the generated edit instructions, we annotate each mined failure with one or more of six semantic error categories---\textbf{Object}, \textbf{Action}, \textbf{Attribute}, \textbf{Count}, \textbf{Relation}, and \textbf{Hallucination}---the same taxonomy used in the edit-instruction prompt (Appendix~\ref{prompt:edit_instr}). The annotation is \textbf{multi-label}: a single false positive may exhibit multiple error types simultaneously, so category percentages sum to more than 100\%.

\subsection{Error Type Distribution}

Table~\ref{tab:error_type_dist} reports the error type distribution for COCO30K across both retrieval directions.

\begin{table}[htbp]
\centering
\small
\begin{tabular}{lcc}
\toprule
Error Type & I2T (\%) & T2I (\%) \\
\midrule
Object Error & \textbf{44.0} & \textbf{48.3} \\
Hallucination & \textbf{36.9} & 0.5 \\
Action Error & 34.1 & 26.6 \\
Attribute Error & 33.7 & 10.7 \\
Count Error & 18.8 & 12.4 \\
Relation Error & 12.1 & 15.8 \\
\midrule
Multi-label density & 179.6 & 114.3 \\
\bottomrule
\end{tabular}
\caption{Error type distribution on COCO30K (multi-label; percentages sum to $>$100\%). Column maxima and the I2T Hallucination rate are shown in \textbf{bold}.}
\label{tab:error_type_dist}
\end{table}

Three observations emerge. First, \textbf{Hallucination exhibits an extreme directional asymmetry}: 36.9\% for I2T versus only 0.5\% for T2I---a roughly 74$\times$ difference. This is intuitive: false-positive captions naturally ``hallucinate'' details not present in the query image (fabricated objects, actions, or attributes), whereas false-positive images are real photographs whose content physically exists and cannot be hallucinated. This stark contrast directly validates RePair's bidirectional independent mining design---I2T editing must frequently remove hallucinated content, while T2I editing never encounters this failure mode. A unified strategy for both directions would severely misallocate synthesis effort.

Second, \textbf{I2T failures are multi-dimensional while T2I failures are more focused}. The multi-label density for I2T (179.6\%, i.e., $\sim$1.8 error types per sample) substantially exceeds T2I (114.3\%, $\sim$1.1 types per sample). False-positive captions tend to err across multiple semantic axes simultaneously (e.g., wrong object \textit{and} wrong attribute \textit{and} hallucinated action), whereas false-positive images typically deviate on only one or two dimensions. This supports RePair's holistic editing strategy: rather than designing type-specific correction pipelines, the LLM adaptively diagnoses and addresses whichever combination of errors is present.

Third, \textbf{Object Error is the most prevalent failure mode in both directions} (I2T: 44.0\%, T2I: 48.3\%), indicating that object-level confusion is CLIP's most common retrieval failure. Count Error (12.4\%--18.8\%) and Relation Error (12.1\%--15.8\%) are comparatively rare, suggesting that CLIP's failures are driven more by entity-level mismatches than by compositional reasoning errors.

\subsection{Editing Quality by Error Type}

To assess whether counterfactual editing is equally effective across error types, we measure the cosine similarity between the original false positive and its edited counterpart in CLIP embedding space for COCO30K T2I (Table~\ref{tab:edit_quality_by_type}). Higher similarity indicates a more minimal, controlled edit.

\begin{table}[htbp]
\centering
\small
\begin{tabular}{lccc}
\toprule
Error Type & Mean & Std & Median \\
\midrule
Attribute Error & \textbf{0.912} & \textbf{0.063} & \textbf{0.932} \\
Relation Error & 0.882 & 0.086 & 0.904 \\
Action Error & 0.875 & 0.093 & 0.901 \\
Count Error & 0.873 & 0.100 & 0.899 \\
Object Error & 0.860 & 0.104 & 0.891 \\
\bottomrule
\end{tabular}
\caption{Editing quality by error type (COCO30K T2I, cosine similarity between original FP and edited counterpart). Hallucination is omitted due to insufficient T2I samples ($n{=}3$).}
\label{tab:edit_quality_by_type}
\end{table}

Two findings stand out. First, \textbf{editing is effective across all error types}: every category achieves a mean cosine similarity $\geq$0.86 and a median $\geq$0.89, with no systematic failures for any type. This confirms the universality of counterfactual editing---no type-specific editing strategies are needed. Second, \textbf{editing quality correlates with operation granularity}: Attribute edits are the most precise (mean 0.912, lowest std 0.063), as modifying a color or size involves minimal embedding perturbation. Object replacements show the largest drift (mean 0.860, highest std 0.104), consistent with entity substitution being the most ``aggressive'' edit operation. Crucially, even Object edits remain well-controlled---RePair's post-edit selection mechanism ($M{=}3$ candidates, selecting the highest cosine similarity) effectively bounds semantic drift for all error types. Appendix~\ref{app:post_edit_audit} provides a direct human check on the same property: $93.3\%$ of retained edits are judged minimal.

\subsection{Editing Failure Modes}
\label{app:edit_failure_modes}

We classify each of the 60 records ($7.5\%$ of 800) that were not rated as unequivocally valid hard contrasts in Appendix~\ref{app:post_edit_audit} into one of four modes: \textbf{semantic mismatch} (the edit does not fully satisfy the query, e.g.\ a count error or an incomplete object replacement), \textbf{scaffold drift} (the edit rebuilds the scene rather than changing the residual), \textbf{artifact} (a visible generation artifact), and \textbf{ambiguous FP} (the original false positive is debatably correct).

\begin{table}[htbp]
\centering
\small
\setlength{\tabcolsep}{4pt}
\begin{tabular}{lccccc}
\toprule
\multirow{2}{*}{Failure Mode} & \multicolumn{2}{c}{COCO30K} & \multicolumn{2}{c}{Flickr30K} & \multirow{2}{*}{Total} \\
\cmidrule(lr){2-3} \cmidrule(lr){4-5}
 & I2T & T2I & I2T & T2I & \\
\midrule
Semantic mismatch & 2 & 8 & 12 & 12 & \textbf{34} \\
Scaffold drift & 0 & 8 & 0 & 13 & \textbf{21} \\
Artifact & 0 & 0 & 0 & 1 & \textbf{1} \\
Ambiguous FP & 1 & 0 & 3 & 0 & \textbf{4} \\
\midrule
Total & 3 & 16 & 15 & 26 & \textbf{60} \\
\bottomrule
\end{tabular}
\caption{Editing failure modes among the 60 shortfall records from the post-edit verification ($N{=}800$).}
\label{tab:edit_failure_modes}
\end{table}

Three findings follow from Table~\ref{tab:edit_failure_modes}. First, \textbf{scaffold drift is exclusive to T2I} (21 of 21 cases). Text edits never rewrite the scaffold, whereas diffusion-based image editing occasionally regenerates the entire scene instead of the residual. This is the concrete mechanism behind the directional gap in Table~\ref{tab:post_edit_audit} and behind the editor-sensitivity results in Table~\ref{tab:ablation_r1}, where a weaker image editor degrades T2I most. Second, \textbf{semantic mismatch is the largest mode} (34 of 60) and is more frequent on Flickr30K (24 vs.\ 10 on COCO30K), affecting both directions---Flickr30K's more varied scenes and free-form captions make the target semantics harder to realize. Third, \textbf{visible artifacts are rare} (1 of 800): selecting among $M{=}3$ candidates by embedding proximity removes most artifact-laden edits before training. The four ambiguous-FP cases all occur in I2T, where whether a caption is genuinely wrong can itself be a judgment call.

\subsection{Editor Sensitivity}
\label{app:editor_sensitivity}

Table~\ref{tab:ablation_r1} swaps the image editor while holding the rest of the pipeline fixed. Three observations follow.

First, every swap moves T2I more than I2T: $-2.25$ vs.\ $-1.57$ for InstructPix2Pix and $-1.10$ vs.\ $-0.82$ for Flux2-4B. That T2I responds most is expected, since this component edits only images. That I2T responds at all is not: I2T captions are produced by the LLM and are untouched by the swap. The effect transfers through joint bidirectional training---degraded T2I pairs perturb the shared embedding space, which the I2T direction inherits. It is the same coupling that produces the positive cross-direction transfer in Table~\ref{tab:merged_results}, where image-side-only training (RePair-T2I) lifts COCO30K I2T R@1 from $60.54$ to $61.78$.

Second, the spread across editors is large relative to the effect being measured: $2.49$ R@1 points on COCO30K T2I between InstructPix2Pix ($43.86$) and Gemini 2.5 Flash Image ($46.35$), against a RePair-over-Vanilla gain of $1.96$. RePair is editor-agnostic in that it needs no editor-specific tuning, but the ceiling it can reach is set by the editor.

Third, a sufficiently weak editor pushes the T2I direction below the untuned baseline: InstructPix2Pix reaches $43.86$ against Vanilla's $44.15$. I2T remains above Vanilla for all three editors. Appendix~\ref{app:edit_failure_modes} identifies the mechanism---scaffold drift, the edit regenerating the scene rather than the residual, occurs only in T2I---so a weaker editor degrades exactly the property that the \textsc{Minimality} principle relies on.

\section{Multi-round Failure-conditioned Synthesis}
\label{app:multi_round}

RePair supports multi-round synthesis as a self-bootstrapping error-mining loop: at each iteration, we re-run retrieval with the current checkpoint, harvest newly exposed failures, and synthesize targeted counterfactual hard samples around these failures for the next training stage. We evaluate this iterative process on \textbf{COCO30K}. As shown in Table~\ref{tab:multi_iter_synth_r1_subdelta}, this iterative densification consistently improves retrieval in both directions, yielding monotonic Recall@1 gains across iterations. The key takeaway is that even after one refinement round, the model still contains structured, mineable failure modes; re-mining under a stronger model surfaces ``harder'' near-miss errors that were previously masked, and converting them into supervision continues to calibrate the confusion boundary.

\begin{table}[htbp]
\centering
\resizebox{\columnwidth}{!}{%
\begin{tabular}{lcccc}
\toprule
Iteration
& $N_{\mathrm{I2T}}$
& $N_{\mathrm{T2I}}$
& $\mathrm{R@1}_{\mathrm{I2T}}$
& $\mathrm{R@1}_{\mathrm{T2I}}$ \\
\midrule
Iter0
& $47.0\mathrm{k}$
& $60.0\mathrm{k}$
& $62.64$
& $46.11$ \\

Iter1
& $58.1\mathrm{k}_{\text{\scriptsize +11.1k}}$
& $81.7\mathrm{k}_{\text{\scriptsize +21.7k}}$
& $63.02_{\text{\scriptsize \textcolor{ForestGreen}{+0.38}}}$
& $46.32_{\text{\scriptsize \textcolor{ForestGreen}{+0.21}}}$ \\

Iter2
& $62.3\mathrm{k}_{\text{\scriptsize +4.2k}}$
& $92.3\mathrm{k}_{\text{\scriptsize +10.6k}}$
& $63.11_{\text{\scriptsize \textcolor{ForestGreen}{+0.09}}}$
& $46.40_{\text{\scriptsize \textcolor{ForestGreen}{+0.08}}}$ \\
\bottomrule
\end{tabular}%
}
\caption{Multi-round synthesis: Recall@1 evolution on I2T and T2I retrieval.}
\vspace{-5pt}
\label{tab:multi_iter_synth_r1_subdelta}
\end{table}

At the same time, the process exhibits a clear convergence pattern with diminishing marginal returns. The incremental R@1 gains shrink markedly from Iter0$\to$1 to Iter1$\to$2, while the volume of newly synthesized samples also drops. This is consistent with an ``error-space contraction'' effect: as training corrects dominant failures, remaining errors become rarer and harder, so further synthesis yields limited additional benefit. Practically, this suggests a simple stopping criterion---terminate when the marginal improvement falls below a threshold---balancing compute cost against saturated returns.

\onecolumn
\section{Prompt Templates}

This section presents the prompt templates used in our RePair synthesis pipeline. The pipeline consists of three stages, each with task-specific variations for I2T (Image-to-Text) and T2I (Text-to-Image) synthesis.

\begin{tcolorbox}[
    width=\textwidth, breakable,
    colback=purple!10!white,
    colframe=purple!75!black,
    title={Prompt for Quality Control},
    fonttitle=\bfseries\color{white},
    arc=2mm, boxrule=1pt
]
\small
\textbf{System Prompt:}
\begin{verbatim}
You are a strict Data Quality Assurance Expert for a multimodal dataset.
Your task is to:
1. Verify if the provided Ground Truth (GT) Image and Text correspond
   to each other.
2. Determine if a "False Positive" (FP) candidate is already TOO SIMILAR
   to the Query, making it unsuitable for an editing task.

Ignore minor missing details. Focus on MAJOR mismatches (e.g.,
wrong gender, wrong main object, completely different scene).

Return your analysis in valid JSON format.
\end{verbatim}

\vspace{\medskipamount}

\textbf{User Template (I2T Task):}
\begin{verbatim}
Image Query: [Provided Image]
Ground Truth Caption: "{gt_caption}"
False Positive Caption: "{fp_caption}"

Step 1: Does the GT caption accurately describe the image?
Step 2: Is the FP caption already a sufficiently good description
        of the image?

Return JSON: 
{
    "gt_valid": true/false, 
    "gt_reason": "explanation",
    "fp_too_similar": true/false, 
    "fp_reason": "explanation"
}
\end{verbatim}

\vspace{\medskipamount}

\textbf{User Template (T2I Task):}
\begin{verbatim}
Text Query: "{query_text}"
Ground Truth Image: [Provided Image]
False Positive Image: [Provided Image]

Step 1: Does the text accurately describe the GT image?
Step 2: Is the FP image already a sufficiently good result for the query?

Return JSON: 
{
    "gt_valid": true/false, 
    "gt_reason": "explanation",
    "fp_too_similar": true/false, 
    "fp_reason": "explanation"
}
\end{verbatim}
\label{prompt:qc}
\end{tcolorbox}

\begin{tcolorbox}[
    width=\textwidth, breakable,
    colback=teal!10!white,
    colframe=teal!75!black,
    title={Prompt for Edit Instruction Generation},
    fonttitle=\bfseries\color{white},
    arc=2mm, boxrule=1pt
]
\small
\textbf{Shared System Context:}
\begin{verbatim}
You are an expert Visual-Semantic Auditor / Failure Analyst.
Your task is to generate a structured editing instruction that corrects a 
false-positive result based on a query.
\end{verbatim}

\vspace{\medskipamount}

\textbf{I2T Specific Template (Image Query $\rightarrow$ Text Edit):}
\begin{verbatim}
### Inputs:
- Query Image: [The image is attached below]
- False Positive Caption: "{fp_caption}"

### Task:
Generate a **single string** value for the key "edit_instruction" that strictly follows:
`[Error Type]: Diagnosis of the visual discrepancy. Therefore, specific editing command.`

### Constraints:
1. **Error Types**: Choose strictly from [Object Error, Action Error, Attribute Error, 
   Count Error, Relation Error, Hallucination].
2. **Action**: Provide a minimal "Replace A with B" or "Remove X" command.

### Examples:
Case 1 (Object Error):
- Image: A photo of a **cat** sleeping on a rug.
- Caption: "A dog sleeping on a rug."
- Output: {{"edit_instruction": "[Object Error]: The image clearly shows a cat, 
  not a dog. Therefore, replace 'dog' with 'cat'."}}

Case 2 (Action Error):
- Image: A man **running** to catch a bus.
- Caption: "A man waiting for a bus."
- Output: {{"edit_instruction": "[Action Error]: The man is depicted in motion running. 
  Therefore, change 'waiting for' to 'running to catch'."}}

... [Additional examples for Attribute and Combined errors]
\end{verbatim}

\vspace{\medskipamount}

\textbf{T2I Specific Template (Text Query $\rightarrow$ Image Edit):}
\begin{verbatim}
- User Query: "{query_text}"
- Incorrect Image (False Positive): [Image is attached]

Based on your analysis, provide a clear, step-by-step instruction for an AI image 
editing tool to modify the attached image so that it accurately represents the query.

The instruction should:
- Be specific about what needs to change
- Describe the subject modifications needed
- Maintain overall image style where possible

Return a JSON object with a single key: "edit_instruction".
Example: {{"edit_instruction": "Change the dog to a cat, and modify the running 
pose to a sitting position"}}
\end{verbatim}
\label{prompt:edit_instr}
\end{tcolorbox}

\begin{tcolorbox}[
    width=\textwidth, breakable,
    colback=orange!10!white,
    colframe=orange!75!black,
    title={Prompt for Synthesis Editing},
    fonttitle=\bfseries\color{white},
    arc=2mm, boxrule=1pt
]
\small

\textbf{I2T: Caption Editing Prompt}
\begin{verbatim}
You are an expert caption editor.
You are given:
1. An original caption that was retrieved but is not fully correct.
2. An editing instruction that explains how the caption should be modified.

Original caption: "{original_caption}"
Editing instruction: "{edit_instruction}"

Your task:
- Rewrite the caption so that it becomes accurate and faithful to the image, strictly 
  following the editing instruction.
- The style should be natural, concise, and fluent English.
- DO NOT return JSON. Only output the final corrected caption as plain text.
\end{verbatim}

\vspace{\medskipamount}
\hrule
\vspace{\medskipamount}

\textbf{T2I: Image Editing Prompt}
\begin{verbatim}
Instruction:
{edit_instruction}

Use the provided image as reference.
Generate an edited image that follows the instruction while preserving 
overall style and quality.
\end{verbatim}
\label{prompt:synthesis_edit}
\end{tcolorbox}